\documentclass[utf8]{frontiersSCNS} 

\usepackage{url,hyperref,lineno,microtype,subcaption,array}
\usepackage[onehalfspacing]{setspace}
\usepackage{scrextend}

\usepackage{soul}
\usepackage{xargs}                      
\usepackage[colorinlistoftodos,prependcaption,textsize=tiny]{todonotes}
\usepackage{xspace}
\usepackage{xcolor}

\newcommand\red[1]{{\color{red}#1}}

\def\keyFont{\fontsize{8}{11}\helveticabold }
\def\firstAuthorLast{Ognibene {et~al.}} 
\def\Authors{Dimitri Ognibene\,$^{1,2,*}$, Rodrigo Wilkens\,$^{9}$, Davide Taibi\,$^{3}$, Davinia Hernández-Leo\,$^{4}$, Udo Kruschwitz\,$^{5}$, Gregor Donabauer\,$^{5}$, Emily Theophilou\,$^{4}$, Rene Alejandro Lobo\,$^{4}$, J. Roberto S\'{a}nchez-Reina\,$^{4}$, Lidia Scifo\,$^{3}$, Francesco Lomonaco\,$^{1}$, Sathya Bursic\,$^{1}$, Veronica Schwarze\,$^{6}$, Johanna B\:{o}rsting\,$^{6}$, 
 Ulrich Hoppe\,$^{6}$, Farbod Aprin\,$^{6}$, Nils Malzahn\,$^{7}$ and Sabrina Eimler\,$^{6}$}

\def\Address{$^{1}$University of Milano-Bicocca, Milan, Italy \\
$^{2}$School of Computer Science and Electronic Engineering, Faculty of Science and Health, University of Essex, United Kingdom \\
$^{3}$Institute for Education Technology, National Research Council of Italy, Palermo, Italy \\
$^{4}$Pompeu Fabra University, Barcelona, Spain \\
$^{5}$University of Regensburg, Regensburg, Germany \\
$^{6}$Ruhr University Bochum, Germany \\
$^{7}$Fakult\:{a}at f\:{u}r Ingenieurwissenschaften,Universit\:{a}t Duisburg-Essen, Germany \\
$^{8}$Cental, ILC, UCLouvain}
\def\corrAuthor{Corresponding Author}

\def\corrEmail{\red{dimitri.ognibene@unimib.it}}

\begin{document}
\onecolumn
\firstpage{1}

\title[Social Media Companion for Collective Well-Being]{Challenging Social Media Threats using Collective Well-Being-aware Recommendation Algorithms and an Educational Virtual Companion} 

\author[\firstAuthorLast ]{\Authors} 
\address{} 
\correspondance{} 

\extraAuth{}

\maketitle

\begin{abstract}

Social media have become an integral part of our lives, expanding our interlinking capabilities to new levels. There is plenty to be said about their positive effects. On the other hand, however, some serious negative implications of social media have  been repeatedly highlighted in recent years, pointing at various threats to society and its more vulnerable members, such as  teenagers, in particular, ranging from much-discussed problems such as digital addiction and polarization to manipulative influences of algorithms and further to more teenager-specific issues (e.g. body stereotyping). 
The impact  of  social media  -- both at an individual and societal level -- is characterized by the complex interplay between the users' interactions and the intelligent components of the platform. Thus users' understanding of  social media mechanisms plays a determinant role.
We thus propose a theoretical framework based on an adaptive “\emph{Social Media Virtual Companion}” for educating and supporting an entire community, teenage students, to interact in social media environments in order to achieve desirable conditions, defined in terms of a community-specific and participatory designed measure of Collective Well-Being (CWB). 
This Companion combines automatic processing with expert intervention and guidance. The virtual Companion will be powered by a \emph{Recommender System} (\emph{CWB-RS}) that will optimize a \emph{CWB} metric instead of engagement or platform profit, which currently largely drives recommender systems  thereby disregarding any societal collateral effect. 
CWB-RS will optimize CWB both in the short term by balancing the level of social media threats the users are exposed to, and in the long term by adopting an \emph{Intelligent Tutor System} role and enabling adaptive and personalized sequencing of playful learning activities. 
We put an emphasis on \emph{experts} and \emph{educators} in the \emph{educationally managed social media community} of the Companion. They play five key roles: (a) use the Companion in classroom-based educational activities; (b) guide the definition of the CWB; (c) provide a hierarchical structure of learning strategies, objectives and activities that will support and contain the adaptive sequencing algorithms of the CWB-RS based on hierarchical reinforcement learning; (d) act as moderators of direct conflicts between the members of the community; and, finally, (e) monitor and address ethical and educational issues that are beyond the intelligent agent’s competence and control. 
This framework offers a possible approach to understanding how to design social media systems and embedded educational interventions that favour a more healthy and positive society.
Preliminary results on the performance of the Companion's components and studies of the educational and psychological underlying principles are presented.
\tiny
 \keyFont{ \section{Keywords:} \red{keyword, keyword, keyword, keyword, keyword, keyword, keyword, keyword}} 
\end{abstract}

\section{Introduction} \label{sec:intro}
Social media (SM) have become an integral part of our everyday lives. Looking at the field more broadly, the freedom to post whatever someone judges useful has been described as nothing less than a shift in the communication paradigm \citep{Baeza-Yates99}, or in other words, \emph{the freedom to publish} marks the birth of a new era altogether \citep{Baeza-Yates10Modern}. There is  ample evidence of positive effects of SM  that goes beyond just-in-time connectivity with a network of friends and like-minded people, including, but not limited to, improved relationship maintenance \citep{ellison2014cultivating}; increased intimacy \citep{jiang2011disclosure}; reduced loneliness \citep{khosravi2016impact,ryan2017social} and reduced depression \citep{grieve2013face}.  
%
It has become a highly accessible and increasingly popular means of sharing content and immediately re-sharing  others' content. Supported by personalizing recommendation algorithms, which suggest content and contacts, SM allows information of any quality to spread at an exponentially faster rate than the traditional ``word of mouth'' \citep{murthy2012towards,Webb2016}. 
%
However, far from creating a global space for mutual understanding, truthful, and objective information, the large-scale growth of SM has also fostered negative social phenomena, e.g. (cyber)bullying to pick just one ~\citep{cowie2013cyberbullying,Mladenovic20Cyber}, that only existed on a limited scale and slow pace before the digital revolution. These issues are escalated by impulsive, alienating and excessive usage that can be associated with digital addiction 
\citep{almourad2020defining}. These phenomena, enabled by the rapid  spread of information on SM 
can affect the well-being of more vulnerable members of our society, such as teenagers, in particular \citep{talwar2014adolescents,gao2020mental,ozimek2017materialists}. 
Ever since the Cambridge Analytica scandal \citep{Isaak18User}, we have become more sensitive to the negative implications of social media. 
One might go as far as to suggest that SM may have become so dangerous that we would be in a better place without them, but that is clearly an unrealistic idea.

It can be argued that the impact of online experience, especially in SM, intrinsically depends on the mutual attitudes and interactions between the members of the community \citep{jones2016defining} and  their interplay with the intelligent
components of the platforms. 
This calls for a holistic approach that on one side provides \emph{educational interventions}  supporting users in understanding the impact of their actions on the experience of the other members of the community \citep{jones2016defining,xu2019social,helmeto_2022_1} and their role in the \emph{Collective well-being of their social media community (CWB)} \citep{roy2018collective,ahn2013social,allcott2020welfare}. CWB operationally combines the different aspects of what a community considers its ``desirable condition'' , while also crucially considering individual differences and conflicting interests.
Moreover, the lack of users' \emph{`new media literacy'} \citep{scolari2018transmedia} (i.e. understanding of social media mechanisms) has a strong role in escalating SM threats. For example, a study with middle-school students found that more than 80\% of them believed that the ``sponsored content'' articles shown to them were true stories \citep{wineburg2016evaluating}. On the other side a multifaceted approach needs to provide technological support to reduce the strain cognitive resources of social media users. 
An important question is how this can be realized considering the complexity of the involved phenomena, the diverse attitudes and interests of the users, the cost of an intervention with a coverage and impact comparable to that of social media.

With this motivation, in this paper, we articulate a framework for educating teenagers in their interaction with SM and synergetically improve and support their experience based on a  ``\emph{Social Media Virtual Companion}''. 
Inside an external SM platform, it will create an \emph{educationally managed social media community} where playful learning activities and healthy content will be  integrated into participants' SM experience. 
Educational goals and interventions will be designed by experts and educators, e.g. to raise awareness about potential threats and to show alternative healthy interactions. 
To select the most suitable content and effective interventions based on experts' and educators' designs, the companion will incorporate functions of an \emph{Intelligent Tutor System} (ITS). 

\begin{figure}[h] 
\caption{The virtual \emph{ Social Media Companion}  enables continue educational and interaction support for 
a community of students with the involvement of educators. This generates an  \emph{Educationally Managed Social Media Community} whose   Collective Well-Being is actively improved by the CWB-RS powering the Companion under the guidance of the educational objectives and strategies provided by the educators.}
\centering
\includegraphics[width=1.0\textwidth]{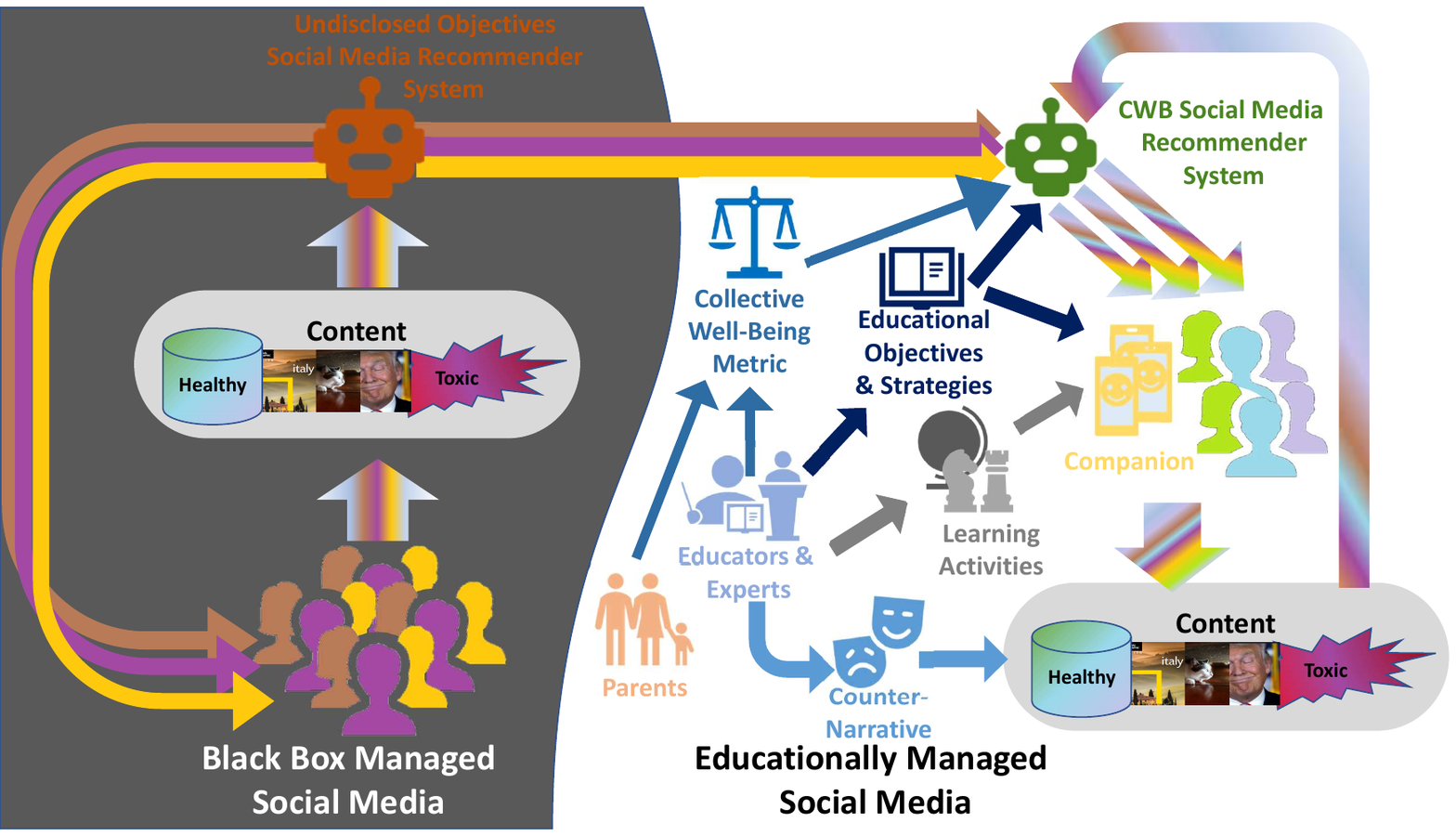}
\label{img:platform}
\end{figure}

Due to the cognitively burdening and overloading information flow of the current SM platform (see sec. \ref{sec:cog_emo_threats} and \citep{weng2012competition, kramer2014experimental, almourad2020defining, Lee2019a}), the Companion will also have to balance and ignore engagement-driven external platform recommendations to target for a fairer and healthier objective \citep{rastegarpanah2019fighting}. 
The community of users of the SM platform is both the producer and consumer of SM content. We affirm that the objective pursued by SM  algorithms  should be closer to the community's needs than those of the SM platform. 
As the CWB reflects the global impact of MS on the condition of the individual and the community, we propose that a suitable objective is a measure that formalizes community-specific and participatory designed CWB expanded with student-specific educational objectives. 
This  shifts to a \emph{CWB} metric evaluated directly on the Companion and used as an optimization target by its integrated recommendation engine (\emph{CWB-RS)}, which will allow the support and  educational management of the local social media community (see fig. \ref{img:platform}).

This framework can be seen as a top-down vision that combines education and technology complements, integrates and helps to balance the diverse efforts targeting specific SM issues. We think that social media phenomena’ complexity, their intrinsically interlinked nature, and their impact on our society demand the production and discussion of such overarching views in the scientific community. Furthermore, our framework, by proposing educational social media that can be separated or linked with the main  social media, would improve the problems that arise from platform-enforced restrictions that hinder experimentation and analysis, especially in extended longitudinal studies. The data collected could be the first step to enabling the definition of adequate regulations and revising SM platform designs to improve their impact on society.

In the next section, a concise overview of the SM threats is presented.
In Section \ref{sec:companion}, we present the educational Companion approach for the increase of digital literacy and the enhancement of CWB. 
 In Section \ref{sec:def_cwb_metric}, we discuss 
the CWB metrics.
The CWB-RS 
is presented in Section \ref{sec:cwb-rs} while, in Section \ref{sec:use_case}, we present a use case exemplifying the interaction of the SM users with the Companion and CWB-RS.  In section \ref{studiesresults} we present the current advances of this line of research.

\section{Social Media Threats} \label{sec:sm_threads}

With the advent of social media, the speed and number of interactions escalated  beyond users'  ability to monitor and understand their impact. This resulted in challenging threats with a broad range and variability over time,  compounded by crucial ethical and practical issues, like preserving freedom of speech and allowing users to be collectively satisfied while dealing with the conflicts generated by their different opinions and contrasting interests. These are magnified by the complex dynamics of information on social media due to the interaction between myriads of users and intelligent artificial systems. 

Critical cases are the pervasive diffusion of fake news and biased content and the growing trend of hate practices. Indeed, hate propagators were among the early adopters of the Internet~\citep{doi:10.1111/j.1530-2415.2003.00013.x,schafer2002spinning,1a228f7d7bd840028c8ec3f29341fce1}. Even though SM platforms presenting policies against hate speech and discrimination, these new media have been shown to be  powerful tools to reach new audiences and   spread racist propaganda and incite violence offline.
%
This gave rise to concern of several human rights associations about the platforms' usage to spread all forms of discrimination \footnote{Simon Wiesenthal Center: \url{http://www.digitalhate.net}, Online Hate and Harassment Report: The American Experience 2020: \url{https://www.adl.org/online-hate-2020}}~\citep{chris2012extremism,bliuc2018online}.

 The social media  threats can be broadly classified into three categories: 1.~content; 2.~algorithmic; network, and attacks; and 3.~dynamics.  However, sharply  separating these types of threats is not trivial as they strongly interact and mutually reinforce while often leveraging on several cognitive aspects and limits of the users.
 In the rest of this section, we briefly discuss SM threats, and we present in Table~\ref{tab:sm_threats} a list of examples per category. We focus on threats that specifically affect the vulnerable population of teenagers and the related threats, such as bullying \citep{fulantelli2022cyberbullying,talwar2014adolescents,Mladenovic2021}, addiction \citep{tariq2012impact,shensa2017problematic}, body stereotypes, and others
 \citep{mcandrew2012does,clarke2009early,ozimek2017materialists}.
 
\subsection{Content-Based Social Media Threats}
The content-based threats are common to classical media, but specific issues 
 thrive on the web and social media in particular. Examples of content-based threats include 
toxic content \citep{Kozyreva2020Citizens},
fake news/disinformation \citep{de2018multi},
beauty stereotypes \citep{verrastro2020fear},    
and bullying \citep{grigg2010cyber}. 
%
Given the importance of these threats, various research is focused on the development of dedicated detection systems as discussed in Section \ref{sec:ThreatDetection}.

\subsection{Algorithmic Social Media Threats}\label{sec:alg_threat}
The SM algorithms  may create additional threats.
For example, the selective exposure of digital media users to news sources  \citep{schmidt2017anatomy}, risks creating a permanent distorting state of isolation from different ideas and perspectives, i.e.    `filter bubbles’ \citep{nikolov2015measuring,https://doi.org/10.1111/bjso.12286},  and form closed-group polarised social structures, i.e. `echo chambers' \citep{ del2016spreading, gillani2018me}. 
Another undesired network condition is gerrymandering \citep{stewart2019information}, where users are exposed to unbalanced neighbourhood configurations. 

\subsection{Social Media Dynamics induced Threats}\label{sec:dyn_threat}
The social media dynamics induced by the extended and fast-paced interaction between their algorithms, common  social tendencies, and stakeholders' interests may also be a source of threats \citep{anderson2012media,milano2021ethical}.
 These factors may escalate the acceptance of toxic beliefs ~\citep{neubaum2017opinion, stewart2019information}, make social media users' opinions susceptible to phenomena such as the diffusion of hateful content, and induce violent outbreaks of fake news on a large scale~\citep{del2016spreading,Webb2016}.  

\subsection{Social Media Cognitive and Socio-emotional  Threats}\label{sec:cog_emo_threats}
While many studies that analyse the mechanisms of content propagation in social media exist, 
how to model the effects of users' emotional and cognitive states or traits on the propagating malicious content is unclear, especially  in light of the significant contribution of their cognitive limits \citep{weng2012competition,pennycook2018falls, Allcott2017}. 
Important cognitive factors are users' limited attention and error-prone information processing \citep{weng2012competition} that may be worsened by the emotional features of the messages \citep{kramer2014experimental,brady2017emotion}. Moreover, the lack of non-verbal communication and limited social presence \citep{KristaR.Mehari2014, gunawardena1995social, rourke1999assessing} often exasperates carelessness and misbehaviours, as the users perceive themselves as anonymous \citep{diener1980deindividuation, postmes1998deindividuation}, do not feel judged or exposed \citep{whittaker2015cyberbullying} and deindividualize themselves and other users \citep{lowry2016adults}.

Over time, users' behaviours can deteriorate and show highly impulsive and addictive traits \citep{kuss2011online}. Indeed, social media 
usage presents many neurocognitive characteristics (e.g. the presence of impulsivity)  typical of more established forms of pharmacological and behavioural addictions \citep{Lee2019a}.  This recently recognised threat, named \emph{Digital Addiction (DA)} \citep{almourad2020defining,pmid26394521,Lavenia2012}, has several harmful consequences, such as unconscious and hasty actions \citep{ali2015emerging, alrobai2016online}. Some of them are especially relevant for teenagers affecting their school performance and mood \citep{aboujaoude2006potential}. In the last few years, it emerged that recognising addiction to social media cannot be based only on the ``connection time" criterion but also on how people behave \citep{TAYMUR2016532,10.3389/fpsyg.2018.00558}. Like in the other behavioural addictions, a crucial role may be played by the environment structure \citep{OGNIBENE2019269, kurth2009temporal}, more than by biochemical failures of the decision system \citep{lim2019impairments}. Indeed, many, if not all, aspects of social media environments are under the control of the 
recommender systems, which may help reduce the condition with specific strategies, such as higher delays for more impulsive users as well as detecting and curbing its triggers, e.g. feelings of Fear of Missing Out 
\citep{Alutaybi2019}.
\subsection{Limited social media literacy}
Finally, the lack of digital literacy, common among teenagers\citep{doi:10.1080/17439884.2013.783597},  can strongly contribute to other threats escalation, for example by favouring the spread of content-based threats and engaging in toxic dynamics  \citep{wineburg2016evaluating}. Teenagers also  show  over-reliance on algorithmic recommendations and a lack of awareness of the unwitting use of toxic content. Thus reducing their ability to make choices
and increasingly deviating toward dangerous behaviours \citep{doi:10.1177/0743915619858057,doi:10.1509/jppm.15.020}.

This diverse set of phenomena and threats, the latter in particular, motivates our educational approach combining educational methods to rise digital citizenship and new median literacy while supporting the user with a smart companion that can also counter the cognitive burden of interacting with social media.


\section{Educational Social Media Companion}
\label{sec:companion}
Social media have been shown to contribute to our collective well-being enhancing our levels of social connectivity. However, our well-being, and in particular  teenagers' one, is  vulnerable to social media threats, such as  exposure to  many types of unwanted or toxic content \citep{Costello2019,Mladenovic20Cyber}.
%
Increasing social media users' digital literacy \citep{Fedorov2015} and citizenship \citep{xu2019social, jones2016defining} may counter most SM threats  that thrive due to users' lack of awareness and over-reliance on algorithmic recommendations    \citep{doi:10.1177/0743915619858057,doi:10.1509/jppm.15.020, doi:10.1080/17439884.2013.783597}. 

The traditional media literacy approaches were based on the idea that media had adverse effects on children. 
Therefore, it was necessary to ``immunize'' young people so they can resist such  negative influence. 
As the media ecosystem evolved, so did media literacy. 
It soon included a paradigm shift towards education and risk prevention concerning the web, video games, social networks and mobile devices. Recently, new concepts have been developed to name these new forms of literacy, from ``digital literacy'' or ``digital citizenship''  to ``new media literacy'' \citep{xu2019social,scolari2018transmedia}. 
With the  objective of contrasting social media threats, several  countries have introduced  educational initiatives to increase the awareness of students with respect to the detection of fake news and misleading information on the web\footnote{Retrieved from here: https://www.bbc.co.uk/programmes/articles/4fRwvHcfr5hYMMltFqvP6qF/help-your-students-spot-false-news BBC, (UK), https://literacytrust.org.uk/programmes/news-wise/ NewsWise (UK)}. 
%
Still, due to their limited duration and their high costs compared to purely entertaining use of social media, the  effects of these programs may be limited.

We propose a framework based on a virtual \emph{Educational Social Media Companion}  that   enables continued, both in the classroom  and outside,  educational and interaction support for a community of learners, creating  an  \emph{Educationally Managed Social Media Community} aimed at improving users' new media literacy and social media experience.
Through  companion  support, the students can safely learn by doing  how to deal with social media content, leveraging the positive aspects and counteracting the inherent threats.
%
The relation between those elements is shown in Figure \ref{img:platform}.

While previous educational attempts have focused on literacy activities mainly about \textit{external} threats,  
improving the impact of social media on our society is challenging essentially because the  interactions between users determine the quality and consequences of their experience. 
Rising awareness about  the effects of own actions on the community members' experience and the importance of  performing healthy interactions to realize a desirable condition  notwithstanding the anonymity \citep{schlesinger2017situated,peddinti2014internet} and deindividuation  that  social media may foster \citep{postmes1998deindividuation,diener1980deindividuation,lowry2016adults} is central in the presented educational endeavour. 

We propose that the educationally managed communities participate in the description of a shared vision of  a ``desirable social media community''  in terms of  an operational \textbf{Collective Well-Being  (CWB)} definition specific for their community (see sec. \ref{sec:def_cwb_metric}).
This will support the coherent formulation of community  regulations, objectives and educational activities that  involve several ethical issues entailing the definition of boundaries and trade-offs to own personal behaviour online (see Section \ref{sec:challenges_defcwb}), such  as enabling  collective satisfaction and  preserving the right to free speech \citep{Webb2016}   while facing the conflicts generated by users' different attitudes, opinions, personal history,  and conflicting interests. A formalization of the  CWB informs the CWB-RS, the companion recommender system aimed at recommending educational  activities and content while balancing the recommendation incoming from  the external social media platforms to improve the community's collective well-being, see sec\ref{sec:cwb-rs}.

\subsection{An Educationally Managed Social Media Community} \label{educational_SMcommunity}

The Companion safeguards teens' interactions on social media and implements \emph{playful adaptive educational strategies} to engage and scaffold them considering personalized \emph{educational needs and objectives}. 
These strategies comprise \emph{scripted learning designs} \citep{amarasinghe2019data} that informing by the CWB-RS will articulate the behaviour of the Companion presenting teens with the right level of educational scaffolding \citep{Beed1991} through an adaptive, personalized and contextualized sequence of \emph{learning activities} and supported social media interaction -- incorporating behavioural and cognitive interventions (\emph{nudges} and \emph{boosts}) that are grounded in behavioural psychology \citep{Thaler2009Nudge,Hertwig2017Nudging,purohit2020designing}.
Game mechanics based on a \emph{counter-narrative} \citep{davies2016toward} approach will support learning activities related to rising awareness: motivation, perspective taking, external thinking, empathy, and responsibility. 
These narrative scripts pursue collective and individual \emph{engagement} with the Companion, offering motivating challenges and rewards aimed at keeping users' interest even in the presence of non-educational social media platforms \citep{van2011game} while maintaining awareness of the digital addiction threat.
The autonomous capabilities provided by the CWB-RS to the Companion  can be particularly helpful outside of the classroom 
to avoid the cognitive overload, addiction or over-exposure to toxic content that the recommender system of an external, non-educational, social media platform may select. Moreover, they allow achieving a level of availability comparable with that of non-educational social media while reducing the   moderating effort requested from the moderating educators.

\subsubsection{Educators and the companion: a  human in the loop  view}

In our framework, the educators not only use the companion for delivering tailored educational activities in the classroom but, together with the experts, participate  in the moderation and support of the community as well as in the definition of its CWB and related educational strategies, which drive the Companion by informing the CWB-RS.
The educators  oversee the CWB-RS behaviour playing a key “human in the loop” role \citep{zanzotto2019human, nunes2015survey}. 
This alleviates the  complexities faced by the CWB-RS, such as noise in the  estimation of content toxicity (see section \ref{sec:ThreatDetection}), which may also lead to misinterpreting users' needs and possibly exacerbating their condition.
While the CWB-RS will have  implicit moderating behaviours, e.g. reducing the presentation priority of users' confrontational interactions, the educators will have a central role in arbitrating users’ disputes as well as solving the conflicts that may emerge between different components of an `under-construction' CWB measure, such as  between emotional health \citep{roy2018collective} of one user and freedom of speech of another. 
%


\subsubsection{Adopting Behavioural Economics to Support Collective well-being}
This educational effort aims to help users of social media make the right decision and teach them the necessary skills to get to that point. 
Strategies developed in the context of behavioural and cognitive sciences offer a well-founded framework to address this issue. In particular, we consider nudging \citep{Thaler2009Nudge} and boosting \citep{Hertwig2017Nudging} to be two 
paradigms that have both been developed to minimize risk and harm -- and doing this in a way that makes use of behavioural patterns and is as unintrusive as possible.

\begin{figure}[h] 
\caption{\emph{Sketch of Companion User Interface} The Companion  will support the students' interaction with social media by  
 contextualizing the content to increase the students’ awareness and allow them to access a more diverse set of perspectives \citep{bozdag2015breaking} and sources. It also explicitly and visually provides the students with an evaluation of the content's harmfulness \citep{fuhr2018information}. The example shows how a piece of imaginary fake news would be contextualized.}
\centering
\includegraphics[width=1.0\textwidth]{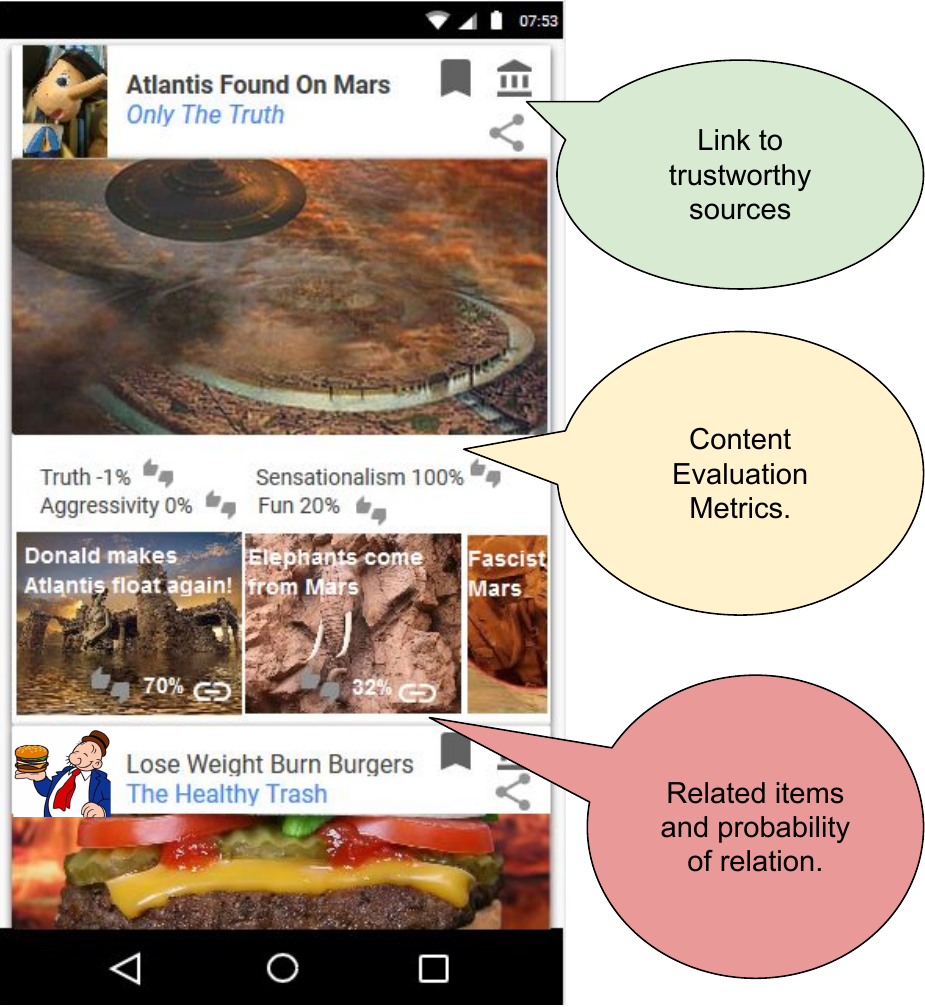}
\label{img:companion_gui}
\end{figure}

Nudging \citep{Thaler2009Nudge} is a behavioural-public-policy approach aiming to push people towards more beneficial decisions through the ``choice architecture'' of people's environment (e.g., default settings). 
%
In the Companion context, such beneficial decisions could be to explore a broad range of different opinions about a specific topic and  check  understandable but scientifically correct  pieces of information. 
In this working example, nudges could be implemented through  a visual layout of the feed that allows easy exploration of such information (see figure \ref{img:companion_gui}). Other forms of nudging are warning lights and information nutrition labels as they offer the potential to reduce harm and risks in web searches, e.g. \citep{Zimmerman2020Towards}. 

The limitation of nudges is that they do not typically teach any competencies, i.e. when a nudge is removed, the user will behave as before (and not have learned anything). This is where boosts come in as an alternative approach. Boosts focus on interventions as an approach to improve people´s competence in making their own choices \citep{Hertwig2017Nudging}. 
In the Companion context, specific educational activities  have been designed  aimed at teaching people skills that help them make healthy decisions, e.g. select/read/trust articles from authoritative resources rather than those reflecting (possibly extreme) individual opinions (see section \ref{studiesresults}). 

The critical 
difference between a boosting and nudging approach is that boosting assumes that people are not merely ``irrational'' and therefore need to be nudged towards better decisions. However, such new competencies can be acquired without too much time and effort and may be hindered by the presence of stress and other sources of reduced cognitive resources. Both approaches nicely fit into the overall approach proposed here. Nudges offer a 
way to push content to users, making them notice. 
Boosting is a particularly promising paradigm to strengthen online users’ competencies and counteract the challenges of the digital world. 
It also appears to be a good  scenario for addressing misinformation and false information, among others. Both paradigms help us educate online users 
rather than imposing rules, restrictions, or suggestions on them.
They 
have massive potential as general pathways to minimize and address harm in the modern online world  \citep{Kozyreva2020Citizens,LorenzSpree2020Behavioural}.

\subsubsection{Educational Activities}
The Companion must also 
provide a satisfying and engaging experience by using \emph{novel hand-defined educational games and activities} based on the interactive counter-narrative concept and educational games. 
SM's entertainment aspect is preserved during the navigation modulated in taking into account CWB, suggesting activities, content, and contacts for the user but managing the exposure to potential threats and addiction. 

The Narrative Scripts help raise users' awareness about SM threats and train the students against them. 
They are  sequences of adaptive learning tasks that provide the right level of educational scaffolding to individuals in developing critical thinking skills, including awareness - perspective taking, motivation, external thinking, empathy, and responsibility) 
by interacting with narratives, counter-narratives, and peers.
These tasks can be different activities, including free-roaming inside the platform, guided roaming following  a narrative, quizzes, playing minigames, or participating in group tasks. 
Different counter-narratives can be triggered depending on students' detected behaviour\citep{hernandez2022narrative}. 

Counter-narrative are used to challenge biased content and discrimination, highlight toxic aspects of messages and attitudes, challenge their assumptions, uncover limits and fallacies, and dismantle associated conspiracy and pseudo-science theories. 

Through a game-oriented setup, the companion bridges the ``us'' versus ``them'' gap that is fostered by hate speech and other expressions of bias (e.g., gendered) and brings forward the positive aspects of an open society and focuses more on ``what we are for'' and less on ``what we are against''.
The users will be informed and  requested to actively and socially contribute to creating and sharing content and material that fosters and supports the idea of an open, unbiased and tolerant society.
Thus, the games can also offer the chance to build connections between the users, which, when isolated, are more vulnerable to online toxic content. 
One approach is to propose periodically specific tests and activities related to each threat, such as \citep{doi:10.1177/0011000010378402}.

A use case scenario is presented in Section \ref{sec:use_case} and the outcomes of several pilot studies that lie the basis for the educational activities are presented in \ref{studiesresults}.


\subsubsection{External and  Internal SM communities separation allows for educational opportunities}
The Companion's location allows it to act as an interface between the educationally managed social media community and the external one. It permits mitigating the effect of external toxic content and offers the opportunity to recreate different interesting experiments about SM phenomena, such as the ones presented in \citep{stewart2019information,backfire1}. 
A controlled environment in which social network dynamics are emulated can be adopted to stimulate students to understand SM mechanisms better, e.g. see \citep{lomonaco2022courage}. Nowadays, the interactions intervening in social media are often mediated by automatic algorithms. Most teenagers ignore these dynamics that heavily influence their content and behaviour when virtually interacting \citep{kuss2013}.
For example, in a classroom, it may expose sub-groups to recommendations with different biases or allow the students to change the recommender parameters \citep{bhargava2019gobo,lomonaco2022courage}. 

\subsubsection{Companion    exposes  social media threats }
The Companion's autonomous mechanisms will support the students in interacting with the social media content both inside 
(as a support learning activities) 
and outside (students' daily social network use) of the classroom. 
The Companion interface exposes its filtering and recommendation algorithms by allowing direct control of their parameters \citep{bhargava2019gobo}. It will contextualize the content to increase the students’ awareness and allow them to access a more diverse set of perspectives \citep{bozdag2015breaking} and sources (see figure \ref{img:companion_gui}). It also explicitly and visually will provide the students with an evaluation of the content's harmfulness \citep{fuhr2018information} 
(see Section~\ref{sec:ThreatDetection}).

\section{Defining a Collective Well-Being Metric for Social Media}  \label{sec:def_cwb_metric}

Social media is an integral part of our everyday lives that is having both negative and positive effects  \citep{wang2014effects,chen2017building}.  
Hence, as positive aspects  rely on the same mechanisms exploited by threats, and because each user's behaviour will affect the other members of the community while  values can differ between communities, it is desirable and necessary to explicitly and collaboratively define shared community principles corresponding to the desired condition of the community.
These community principles will constitute the foundation to define a specific measure of the overall impact of social media in the community  at an individual and a societal level, that is, 
to measure the desirability or \emph{Collective Well-Being (CWB)} of 
a certain condition of the social media community \citep{roy2018collective}. 
These community principles, formalised in the CWB measure, together with an understanding of the virtual and physical social dynamics in the community, should drive the definition of users' behaviour guidelines and connected educational objectives to reach and maintain the community in the desired condition, or in other words, to achieve a high level of CWB.
 A quantitative measure of CWB allows for a more accurate evaluation of the impact of different aspects of the interaction on the community while taking into account the complex and fast dynamics of social media. 
When CWB is estimated directly on the SM  platform it could allow directing its autonomous components, e.g. recommenders, to collaborate in achieving the desired community condition. This would be a  more democratic and transparent objective than the  ones currently pursued by the social media platforms \cite{gorwa2019platform}.
In our framework, it is used to direct the algorithms at the interface between the educationally managed community and the external social media.

\subsection{Research on collective well-being and social media}\label{sec:cwb_lit}
The literature presents several definitions and  measures of well-being \citep{topp20155,gerson2018social}. Some of them were applied in the context of social media to estimate their effects \citep{mitchell2011internet,verduyn2017social,kross2013facebook,chen2017building,wang2014effects} but  mostly considering the single individual with limited consideration for the overarching social aspects  \citep{helliwell2003s}. 

Gross Domestic Product (GDP) has been proposed as an index of the economic well-being of a community\footnote{Retrieved from: https://voxeu.org/article/defence-gdp-measure-wellbeing}. In such contexts, inequality is also an important factor, and it is common practice to use the Gini index to measure it \citep{osberg2017limitations}. 
While the economics view is  difficult to connect to a social media context,  they share similar key issues: which aspects to measure and, above of all, how to compare and aggregate  measures of individuals' well-being  to synthesize that of the whole society \citep{costanza2014development}, even if in this work we consider only the local educational community.

Multidisciplinary notions of CWB extend that of individual well-being to measure a group-level property (construct). They include community members’ individual well-being incorporating diverse domains, such as physical and mental health, often stressing the presence of positive conditions.
They study  which properties of the community affect the  members and how  much each of these properties adds to a comprehensive measure of collective well-being. 
We already stressed the importance of education and educational objectives to support constructive interactions and achieve desirable community conditions, i.e. a high level of well-being. However, education itself is often already part of well-being frameworks \citep{roy2018collective,spratt2017wellbeing,white2007wellbeing,michalos2017education}.  The connection between education and well-being has been analysed from several perspectives. In our framework, the most relevant one is the one defined as \emph{social and emotional literacy} in \citep{spratt2017conceptualising}.

 
 \cite{roy2018collective} present a CWB framework divided into different domains and  comprising health-care and non-health-care-related community factors where the contribution of the latter ones is supported by evidence of their effects on health. This framework can  help to define a checklist for the definition of a community-specific CWB and related measures and indicators. We show in Table \ref{tab:cwb-categories} the properties that may be relevant for education and social media communities for the following reasons:
\begin{itemize}
    \item \emph{Opportunity} domain is related to ``the perceived opportunity to achieve life goals and socioeconomic mobility'' \citep{diener2006measure} as well as the access to education. Social media can be a powerful tool for accessing many opportunities. Feeling in control while using them, instead of just a distraction or worse an addiction, may be an important part of CWB for SM;
    \item \emph{Connectedness} domain is related to the presence of supportive, high-quality, reciprocal relationships with secure attachments. Includes dimensions of social acceptance and social integration that depend on the behaviour of other members of the community \citep{walker2011social};
     \item \emph{Vitality} domain covers many emotional aspects of several individual well-being definitions, such as Fredrickson’s one and Seligman’s model of flourishing \citep{fredrickson2004broaden,seligman2012flourish}.
    However, spillover effects \citep{helliwell2003s} and emotional influence make vitality an important aspect also at a social level;
    \item The threats presented in Section \ref{sec:sm_threads} would impact  negatively the affects component of the \emph{Vitality}  and \emph{Connectedness} domains;
    \item The  \emph{Contribution}  domain relates to community engagement  and related feelings of meaning and purpose. Contribution  can improve other members' experience but may also have negative effects; 
    %
    \item The \emph{Inspiration} domain relates to creativity and lifelong learning, areas where social media have a huge potential.
    \item The  psychosocial \emph{Community} characteristic that is clearly relevant for social media settings:
\begin{addmargin}[2em]{2em}
\textit{``A community with a negative psychosocial environment is one that is segregated and has high levels of perceived discrimination and crime, high levels of social isolation and low community engagement, and low levels of trust in government and fellow citizens.’’} \citep{engel2016older, mair2010neighborhood, klein2013social}.
\end{addmargin}
Community is partially overlapping with the Connectedness and  Contribution domains but describes aspects that are easier to concretely measure in social media networks.
\end{itemize}

While these  formulations of CWB can inspire a guideline to define social media communities' principles and CWB metrics, they must be extended and formalized to better take into account the specific issues and opportunities of SM and in particular, the threats reported in section \ref{tab:sm_threats}. Another important aspect to address is combining contrasting factors or, in other words, formalizing the complex ethical decisions induced by the conflicts and trade-offs that emerge in any social context  \citep{muller2020ethics}.

\subsubsection{Challenges of defining collective well-being for social media} \label{sec:challenges_defcwb}

Defining a CWB metric for SM is an ambitious endeavour  that requires a combined effort of different 
disciplines. 
It would range from political sciences, sociology and psychology over ethical considerations all the way to computer science, machine learning and network theory. 
Besides CWB aspects for physical societies,  the impact of integrated intelligent agents  must also be taken into account in the context of social media, 
as discussed in sections \ref{sec:alg_threat} and  \ref{sec:dyn_threat}. 
A CWB measure for virtual communities has to take into account the conflicts between members as they are frequent and algorithmically augmented.
Therefore, the conflict between the right to freedom of expression,  user satisfaction, and social impact must be stressed more when defining a social media CWB than with physical societies where these factors have slower and better-understood effects and may have  regulations already in place \citep{Webb2016}. 

Conflicts between members' interests pose serious ethical concerns that are out of the scope of this paper and have been the focus of recent research in AI and ethics in different domains \citep{milano2021ethical, cath2018artificial, king2020artificial}. 
When social media are integrated into an educational framework, the problem may be mitigated  by involving educators and experts as moderators. 
We propose that such an educational  setup can also allow initial studies of the implications of a social media platform that aims to improve CWB.

\subsection{Participative definition of social media community principles and CWB factors}
Social media community principles and corresponding CWB factors must be shared by the members of the community.
While research in the field can inform about common social aspects, internationally acknowledged  human rights, or social media-specific phenomena, a community would most likely have the freedom to define tailored principles.
To achieve this human-centred approaches to the participatory design of technology are being explored by the researchers. These approaches involve the stakeholders in the analysis of relevant factors and the co-design of technological solutions. One of the main challenges is bridging the gap between the community members' knowledge and the complexity of cyber-social systems like social media \citep{devito2018people}. An example is a qualitative study to explore adolescents’ representations of social media based on pictorial metaphors, reported in \citep{Sanchez-Reina2022}. The study proposed and analyzed the outcomes of a school project entitled “The Social Media of the future”.  Discourses and visual representations of a total of 168 drawings about their visions for their ideal Social Media tools were analyzed. The results of the analysis pointed out that the relevant CWB factors shared by the adolescents participating in the study were:  care about additive features, transparency in the conflict of interest behind the SM business, also in terms of agency to be able to monitor and control privacy and security facets.

\subsection{Toward the automatic estimation of  collective well-being in social media communities}\label{sec:cwb_cc_cs_ce}
Social media are strongly integrated with information systems that can affordably offer a huge amount of data with a high frequency.
Transforming this data for the estimation of suitable collective well-being measures through machine learning methodologies would open the way to many research and applicative opportunities, such as  autonomous systems that  maximise CWB and avoid current issues induced by profit-based objectives.

Current CWB formulations are not easy to estimate directly using data available  in real-time  on social media, which is necessary to support an autonomous system optimizing CWB.
Moreover, such formulations need to be extended to take into account specific social media issues. 
For example, most of the available formulations of collective well-being focus on positive aspects. 
Nevertheless, the  positive  aspects  (see sec. \ref{sec:cwb_lit} and negative ones (see sec.\ref{sec:sm_threads}) need to be explicitly considered as part of the CWB as they strongly affect social media users and in particular teenagers.

We propose to define a \emph{collective well-being metric for social media} by combining  suitable components of classical CWB and SM threat measures.
The measures of these  components   could be measured  by periodically proposing specific surveys and activities \citep{loughnan2013sexual}. 
However, we propose that additional richer and more transparent measurements  can  be performed by developing intelligent components that analyze users' behaviours. 
In this definition, for each user, event, i.e. content or connection related,  and aspect defined relevant for the CWB three terms are computed: 
\begin{itemize}
    \item  \textbf{CS(aspect, user)} Content Shared measures the aspect-specific value of the content shared by the user;
    \item \textbf{CE(aspect, user)} Content Exposure measures the aspect-specific value of the  content observed by the user;
    \item \textbf{CC(aspect, user a, user b)} Contact Creation measures  the aspect-specific value of new connections based on the participants' CS and CE. 
\end{itemize}

These elements account for the double role of each member of the social media community as  both receivers and producers of 
content.  
In our educational setup, where only the community of interest is in contact with an external social media community,  we distinguish between  ``endogenous'' and ``exogenous'' aspects.
The community can be exposed to threats that are generated outside but a community can also generate such threats inside as part of the interactions in the social medium. 
In this case, the feeds from external sources may be weighted differently.

%
%
%
%
%
While the CS can be seen as a direct expression of the state of the user, it strongly depends on the user’s style of interaction. 
Moreover, only relying on the content shared by users  would induce a substantial delay compared to the moment when a user got actually affected by  observing a piece of content (CE). 
Conversely, the user is exposed to a multitude of diverse inputs   hindering  the interpretation of the overall effect  only from the CE, while the user's reactions (CS) may be more indicative of the most impacting events.
Indeed, current affective state estimators and toxic/positive content detectors can only provide noisy estimations of the current user state and the content quality.
However, the availability of complementary data with higher reliability is limited. 

Once each event is scored for each aspect of interest, it must be decided how to aggregate these terms over users, time, and the different aspects to obtain an estimation of the total CWB of the community. Indeed, the definition of an actual metric following this strategy requires making a number of choices. For example, about the scale for the terms of different aspects considered. 
Regarding aggregation over time, CC, CE and CS values   could be simply   averaged. Other approaches could be considered to take into account the frequency of the events or the  diversity of opinions presented or give more relevance to extreme events, which may be more accurately detected and evaluated. In particular, the value of being exposed to multiple opinions (time-aggregated CE) may be augmented with a measure of diversity (e.g. entropy) \citep{Matakos2020MaxDiv,garimella2017balancing}.

Clearly, the design of the CWB metric presents a number of challenges requiring careful consideration even for  small educational communities that our framework targets. In devising their solutions often the naive approach may at best be ineffective, and at worst exacerbate the issues it was intended to solve.
For example, the aggregation over the aspects dimension may not seem complex when considering the aspects to be independent. In reality, the impact of the various aspects on the users may be interlinked, for example over exposure to content focused on one aspect (e.g. videogames) may lead to overuse of the platform or tire the user who will lose the opportunity to learn about more important content (e.g. social issues).

The most complex  aggregation to design  is over users because it has to balance  the well-being of different individuals and groups of users taking into account their conflicting interactions along different dimensions. 
It is important to consider the different features of each user while respecting privacy constraints. 
For example, vulnerable users are often victims of toxic content but also producers \citep{bronstein2019belief,bessi2016personality,march2019belief}, which affects the CS value.  It is important that they are not isolated  \citep{burrow2017many} and that, at the same time, the toxic content should not be fed to those who could be more affected and instead presented to educators or other community members that have shown constructive reactions to such type of content. This means that the content exposure (CE) should be differently weighted for different community members based on their resilience and that supportive connection creation (CC) should be favoured between people with high  and lower resilience. Still, it is important that resilient members are not overloaded with toxic content and support responsibilities  \citep{steiger2021psychological}.

Apart from the weighting issue issues, another important format issue is the selection of the actual aggregation function across users.
Adopting the naive average  a society where a few radicalized users share extremely hateful content may have a higher CWB
score than one with a number of users sharing content about action movies with slightly violent scenes. 
Another reason why a linear combination of components may not be suitable in the definition of a well-being measure is  that it will simply induce maximizing the terms with positive weights and minimizing terms with negative ones, without allowing a balance. 
For example, if interactions between drastically opposite opinions are considered negative because of possible backfire effects and flames \citep{backfire1}, and interactions between excessively similar opinions are also considered negative because of the echo chambers they may give place, then also interactions between moderately different opinions will have a negative value even when they may lead to a reduced polarization.
%
%
Other aggregation functions may be chosen but it is still difficult to find general solutions. For example, defining the well-being of society as the well-being of the member with lower well-being (i.e. minimum instead of an average) could lead to focusing all the resources on factors that may not be actually changed.

\subsection{Network Measures for Collective well-being on Social Media}
Network-specific measures \citep{rayfield2011connectivity} can be an important part of   an actionable  CWB measure for social media. 
Several threats and well-being-related phenomena are implicitly defined in terms of network measures. 
These measures may also be particularly useful as proxies of future critical conditions without having to execute expensive simulations.  
For example,
\cite{RePEc:eee:phsmap:v:563:y:2021:i:c:s0378437120307895} show 
that  the increase of a network measure of inclusiveness promotes the efficiency and robustness of a society.
\cite{stewart2019information} 
show that an unbalanced network structure may lead to suboptimal collective decisions. 
Effects of positive and negative interactions  at a network level have been studied in \citep{leskovec2010signed}.
Concepts like social influence and homophily \citep{Guo2015,aral2009distinguishing} play an important role in the formation of different network conditions, like segregation, that are crucial for CWB. 
The diversity measures already proposed as part of the CE, CS and CC elements would also contribute to a higher CWB  rating for diversified and integrated communities than polarized and segregated ones.
Other measures viable to characterize user roles, such as centrality and  closeness, can also be used to aggregate the individual users' threat  scores over the network \citep{manouselis2011recommender,drachsler2008identifying}.

\section{An educational Collective Well-Being Recommender System} 
\label{sec:cwb-rs}


Recommendation systems (RSs) are ubiquitous in online activities and are crucial for interacting with the endless sea of information that the Internet and social media present today. In social media platforms, they have introduced the possibility of personalizing  suggestions of both content and connections  based on the use of  user profiles containing also social features \citep{ Eirinaki2018RecommenderSF, heimbach2015value, chen2018survey}. Their goal has been to maximize the  users’ engagement in  activities that  support the platform itself. However, these self-referential objectives  fail to consider  repercussions on users and society, such as digital addiction \citep{almourad2020defining}, filter bubbles \citep{bozdag2015breaking}, disinformation wildfire \citep{Webb2016},  polarization \citep{rastegarpanah2019fighting}, fairness \citep{ranjbar2021fairness,DBLP:journals/corr/abs-1907-13158},  and  other issues discussed in Section \ref{sec:sm_threads}.
To address this, we propose the concept of \emph{Collective Well-Being aware Recommender Systems (CWB-RS)}. The CWB-RS extends social media RS intending to maximize the cumulative long-term \emph{CWB metric} instead of self-referential platform objectives. Compared to previous efforts in dealing with possible negative effects of RSs \citep{rastegarpanah2019fighting,ranjbar2021fairness,DBLP:journals/corr/abs-1907-13158}, the CWBRS takes into account multiple issues and,  to reduce their cumulative impact on society, it adopts longer terms strategies fitting into our  educational framework. 

Integrating  educational objectives aimed at achieving \emph{CWB} in the longer term the CWB-RS will  also have functions  similar to those of a (collective) \emph{Intelligent Tutoring System} \citep{greer2016evaluation}.
RSs have been widely used in educational settings \citep{manouselis2011recommender}, and they are receiving increasing attention due also to the fast growth of MOOC \citep{romero2017educational} and the availability of big data in education \citep{seufert2019pedagogical}.
In educational contexts, recommendations are sequential and functional to achieving  learning goals \citep{tarus2017hybrid}. 
Similarly to the social media context, they have also  employed social information \citep{elghomary2019dynamic, kopeinik2017supporting}.  However, they are usually acting on the content provided  by educators with educational aims, while CWB-RS 
also has to redirect disparate content flowing from external Social Media toward achieving educational objectives.

\begin{figure}[h] 
\caption{\emph{Role of the CWB-RS in the Companion.} CWB-RS will process the \emph{content generated by the users} of the \emph{educationally managed social media} and the \emph{content externally recommended} for them by the RSs of the external social media platform to create new recommendations aimed at maximizing the cumulative long-term \emph{collective well-being metric}. \emph{Content Analyzers and Threat Detectors} will analyze and evaluate the level of threat for each piece of content and other relevant information as the users’ emotional state. This information will be used to: 1) \emph{augment} the information provided to the users by the companion interface; 2) \emph{evaluate} through \emph{predictive models of users’ opinions and reactions} the future effects of different sequences of re-ranking and recommending actions; 3) \emph{select} the re-ranking and recommending actions that resulted in the highest expected cumulative improvement in terms of learning objectives, CWB metrics, agreement with selected educational strategies and user engagement.}
\centering
\includegraphics[width=1.0\textwidth]{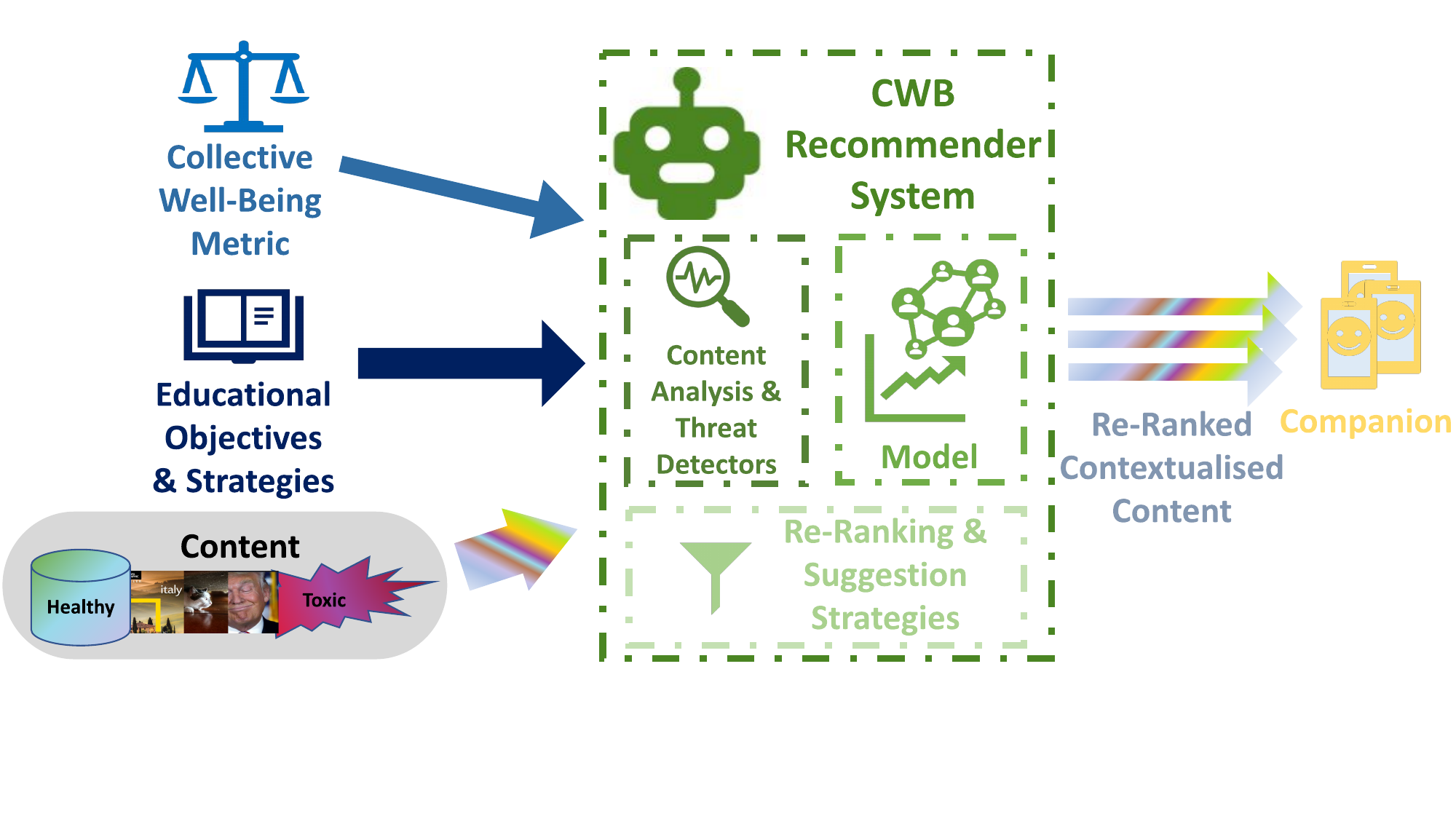}
\label{img:cwbrs_companion}
\end{figure}

As shown in Figure \ref{img:platform}, the CWB-RS creates new recommendations presented through the Companion by processing both the content generated \emph{internally} by the members of the \emph{educationally managed social media} community and the content recommended for them by the RSs of the \emph{external}  platform.
\emph{Content Analyzers and Threat Detectors} (see Figure \ref{img:cwbrs_companion} and sec \ref{sec:ThreatDetection}) will analyze each piece of content to evaluate the level of threat  and other relevant information for the CWB metric, such as the users’ opinions and emotions  (see sec. \ref{sec:cwb_cc_cs_ce}). This information will be used to: 
1) evaluate the current condition of the users;
2) \emph{augment and contextualize} the content provided to the users; 
3) \emph{evaluate}  the future effects of different sequences of content re-rankings and  recommendations through predictive models of users’ conditions; 
4) \emph{select} the  actions that account for the highest expected, long-term, cumulative  CWB metric.   


\subsection{Educational directions for the CWB-RS}\label{sec:educationalobj}
CWB-RS educational objectives are designed by educators and experts (see sec.\ref{sec:companion}).
They  can  be encoded in terms of measures related to specific threats or other well-being variables, such as those extracted by \emph{content analyzers and threat detectors} allowing to easily combine educational and regular CWB objectives  \citep{van2017hybrid}. Different approaches have been proposed to effectively combine and scale multiple terms in  objective functions \citep{marom2018belief,harutyunyan2015expressing}.
These objectives express how much each student: (a) is conscious of his role in other users' well-being, (b) improves his behaviour, and (c) is having a healthy experience. For example, an objective would be `curb obsessive selfies posting' \citep{ridgway2016instagram}, which would act on the content shared (CS) for the aspect `selfies'.
Another example  could be breaking the filter bubbles focused on racist content and helping users hold an unbiased mindset (reduce both CE and CS on the aspect `racism'). In this case, the connected recommendation strategy will be to  provide  content with opposite but not confrontational perspectives  \citep{garimella2017balancing,Matakos2020MaxDiv,bozdag2015breaking}. 
This strategy can be combined with educational games proposing specifically themed challenges, such as finding pictures of achievements performed by people of different ethnicities, suggesting changing the recommender filter parameters directly, or just reducing the racist content presented and substituting it with low harm feeds. 
The CWB-RS can also recommend content to asses the current student's condition \citep{zhou2010solving,kunaver2017diversity} to inform successive personalized interaction.

%
%
Educators and experts will also define interaction strategies specific to each objective \citep{griffith2013policy}. Sketches of \emph{high-level CWB-RS educational strategies} will be hand defined by the educators and experts to choose between the different educational objectives for each student in an effective and contextualized manner. \emph{Lower-level educational strategies} for the CWB-RS comprise hand-defined \emph{learning activities} and \emph{minigames} as well as modulation of the recommendations, for example, showing diverse content as tests to explore students' preferences.

Engagement is an important factor  for both social media platforms \citep{zheng2018drn, wu2017returning} and  educational activities  \citep{sawyer2017balancing}. The CWB-RS must  prevent students from ``dropping out'' \citep{eagle2014modeling,yukselturk2014predicting} and moving to  non-educational social media. In a complementary manner to the game-oriented motivational mechanisms of the Companion \citep{van2011game}, the CWB-RS must therefore preserve a healthy level of engagement \citep{chaouachi2012mental,arroyo2007repairing,mostafavi2017evolution,zou2019} while  avoiding excessive exposure to toxic content as well as any form of addictive use \citep{almourad2020defining,pmid26394521,Lavenia2012}.






\subsection{Challenges in social media RS and CWB-RS}

The realization of effective social media recommendation systems, as reviewed in \citep{Eirinaki2018RecommenderSF,chen2018survey}, presents several challenges that in recent years have brought drastic changes to the field. In particular, some of the biggest challenges are 
the highly diverse information they process (e.g. content, trust, connections),
the complex dynamics of the interactions,
the fast pace of growth of the social graph,
and the enormous amount of multimedia and textual elements to process \citep{ eksombatchai2018pixie, covington2016deep}. 
In the case of the  CWB-RS, the size of the internal social network is limited (i.e. the number of students)
and a big part of the data will come preselected by the external RS, 
thus forming an implicit two stages approach \citep{borisyuk2016casmos,covington2016deep, ma2020off}  with only the second stage in charge of the CWB-RS.
However, the creation of a CWB-RS presents several other theoretical, technical and ethical challenges that are mostly not faced by classical RS.

\subsubsection{Diverse internal and external content}
A first demand for  the CWB-RS is   to combine content defined by the members of the educationally managed social media with recommendations from the external social media. 
While this controlled separation from the external platforms offers the opportunity for novel educational experiences, the heterogeneous nature of signals and structures poses the question of how to combine them. 
This is all conceptually similar to some of the major challenges and opportunities of enterprise and intranet search compared to general web search \citep{Kruschwitz17Searching, Hawking10Enterprise}.

\subsubsection{Social information}
In classical social media RSs, the use of social information is relatively straightforward. For example, connections between users can be interpreted as a cue of similarity between their interests. For a CWB-RS, sharing content based on  social connections may  spread toxic content, however, it can be useful if one of the connected users
has exemplary behaviour. 
Moreover, social network structures affect not only information propagation but also decision and behaviour \citep{stewart2019information}. Thus in CWB-RS, some properties of the structure of the social connection graph of the internal community  may be part of the objective (e.g. CC in sec. \ref{sec:cwb_cc_cs_ce}). 
Still, the recommendation and creation  of connections between diverse groups may sometimes lead to toxic behaviours, e.g. backfiring \citep{backfire1}.

\subsubsection{Lack of direct reference information for the CWB-RS}
Classical RSs  
maximize the users' satisfaction and engagement, usually estimated through accessible proxy measures, such as time of usage or likes. These allow the definition of reference information or teaching signals to improve the RSs behaviour based on the similarity between items or between users' previous selections \citep{Eirinaki2018RecommenderSF,wu2017returning}.
These signals do not inform about the level of CWB or achievement of user-specific educational objectives. 
The CWB-RS needs both to estimate less accessible quantities, such as knowledge acquired or behavioural improvement, and to recommend content taking into account the users' learning trajectories, comprising their current state and assigned objectives. 
Still, these measures  do not easily translate into future recommendations. 
For example, if a recommendation led a student to achieve an educational goal, this does not imply that it would be useful to suggest similar content to the same student again, as it will not provide him with new educational information.  It may still indicate that it is useful to suggest similar content to other students who have to achieve the same goal.



\subsubsection{Temporal aspects and sequence of recommendations}
Classical RSs regard recommending as a static process mainly focusing on ``the immediate feedback and do not consider long term reward'' \citep{liu2018deep,zhao2019deepRLRS}. Instead, to achieve  lasting CWB and the related educational processes, it is necessary to account for the effects of sequences of recommendations. For example, sequencing of lectures, tests, and feedback, is common in most educational strategies.
In addition, a classical RS does not consider the interdependence 
between users’ preferences and the RS recommendations,
which is  crucial to model and counter the filter bubble and echo chamber phenomena. 
Another reason for the CWB-RS to consider a temporal dimension is to enable the use of an accurate dynamic model of the students and the natural variation of their preferences \citep{zeng2016online}.
This 
allows, for example, to prepare the conditions and select the best time for exposure to content aimed at improving students' empathy as well as avoiding wrong conditions, such as those with a high level of user stress, when such content would be ignored or even lead to backfire \citep{backfire1}.

\subsection{CWB-RS adaptation and personalization through Reinforcement Learning}
The 
Reinforcement Learning (RL) paradigm adoption to drive the adaptation and personalization of the CWB-RS behaviour  
\citep{zhao2019deepRLRS,zou2019} 
is a natural solution to the sequential control, lack of supervised teaching signal, and the other   technical issues described above. 
RL-based recommender systems are recently gaining attention in the community \citep{shani2005mdp, liu2018deep, zheng2018drn} 
because of their flexibility,  
 and  the growth of the deep reinforcement learning field \citep{zheng2018drn, mnih2015human}. As suggested in \citep{zhao2019deepRLRS}, RL-based RSs allow solving not only the problem of frequent updates of the user profile, typical of RS in social media, and offer also a precise formulation of the initialization problem in terms of exploitation-exploration 
 \citep{hron2020exploration,iglesias2009reinforcement}. 

From a machine learning perspective, \emph{CWB-RS} educational objectives, learning strategies and activities, can be respectively seen as manually defined rewards, sub-goals, and sub-policies in a Hierarchical Reinforcement Learning (HRL) framework \citep{zhou2019hierarchical,zhou2020improving} which improves its adaptation performance by breaking down the high-level decisions (e.g. the educational objective a student must achieve) and the step-by-step decisions (e.g. which activity or content to show at the moment). 
%
This reduces the  computational costs and amount of data necessary to derive the educational policy and objectives directly from the long-term optimization of the CWB metric \citep{barto2003recent}.

Both classical RL \citep{iglesias2009reinforcement, 
dorcca2013comparing, 
zhou2017towards} and HRL have been used in ITS \citep{zhou2019hierarchical, zhou2020improving} and  RS. 
To our knowledge, this is the first time they are combined. 
While the field of RL-based ITS 
is still young and presents several limits \citep{zawacki2019systematic}, it could address the complex problem of supporting students dealing with the diverse and enormous environment of  social media.
Still, the additional flexibility of RL-based RS comes at the cost of higher complexity,  particularly in terms of training and evaluation setup \citep{henderson2018deep}, as well as deploying in real-world applications \citep{Rotman2020risk, dulac2019challenges}.

\subsubsection{Difficulty of creating CWB-RS datasets}
Reinforcement Learning systems developed to act in real-world conditions are usually pretrained offline on available datasets. Much of the solution quality depends on the similarity between the dataset and the application setting \citep{Rotman2020risk}. 
%
The creation of real-world reinforcement learning datasets most often requires ad-hoc solutions. 

The collection of CWB-RS datasets must take into account the users' profiles, which may be  gathered using a self-reported survey, as in \citet{khwaja2019aligning}, as well as users' neighbourhood  information,  behaviours (e.g. posts) and observations (e.g. recommendations). Mining this information,  however, needs to comply with  privacy and company policies.
%
Additional challenges are presented by the necessity to cover the various reactions that students may have under exposition to combinations of disparate  social media \citep{zhao2019deepRLRS}.
Social media  show a complex interplay between the individual, social, and technological levels of filtering \citep{https://doi.org/10.1111/bjso.12286,gillani2018me}, with  substantial effects on users' behaviours. 
Therefore, one of the strongest challenges 
is washing out the effects of the RS adopted during the data collection, which functioning is usually unknown, enabling the use of the dataset to train a CWB-RS that could propose diverse recommendations and induce  different selections.

\emph{Crowdsourcing}~\citep{boudreau2013using} can be used for large-scale evaluations or for creating datasets under limited  periods~\citep{kittur2008crowdsourcing}. However, special care needs to be taken to ensure the reliability of crowd data~\citep{buhrmester2018evaluation} as the seriousness with which volunteers take their interactions with the system can be limited.
These complexities demand to devise an effective strategy to build a real-world dataset that  considers including the micro-, meso-, and macro-structure, different sources, and modalities.

\emph{Model-Based RL} For the specific setting of the educationally managed social media community, the task is simplified considering  the reduced content variety compared to the external community. Also, while a CWB-RS must be aware of the condition and behaviour of the entire community, this may be factored in terms of the dynamic models of its members. Using different combinations of the same members' models, it could be possible to create different community models that allow a broader set of training conditions for the CWB-RS in simulation. They will also enable online simulations for  estimating the results of a sequence of recommendations (see Figure \ref{img:cwbrs_companion} and \citep{zhao2019model,schrittwieser2020mastering}). The literature on interaction models for social media is extensive. 
\citep{Szabo2010} were  one of the first to show the importance of cognitive and content factors. The models proposed in \citep{Guo2015,he2015hawkestopic} reason simultaneously on the patterns of propagation and the topics. 
Most of these models do not account for user adaptation,  which is crucial in this context. However, the solution could be to adopt generative models of adaptive user behaviours, such as  \citep{OGNIBENE2019269, lindstrom2019computational,Das2014}.
While these studies and many more led to improved forecasting systems, there is a consensus that there are intrinsic problems that limit the predictive power with both sufficient accuracy and anticipation, see for example \citep{cheng2014can}. A significant improvement of baseline algorithms requires  very detailed information about the community \citep{watts2011everything}. However,  the CWB-RS has access to rich information about the educationally managed network. This, together with its limited, size   will improve the efficacy of the  predictive models.

\subsubsection{Risks in the exploration phase of RS based on RL}
Reinforcement learning can  provide online adaptation to conditions that detach from the training set used for offline pretraining. However, this comes with exploration costs that in real environments can pose prohibitive risks \citep{Rotman2020risk}. Even if the CWB-RS is not facing critical safety tasks like those of self-driving systems, repeated sub-optimal recommendations may just reinforce the threats  the Companion is trying to address. 
To alleviate these issues adaptive novelty detection methods \citep{Rotman2020risk} will be implemented in the CWB-RS to recognize  situations far from the agent experience and hand over the control to educators or a safe controller.
Moreover, the HRL paradigm has been adopted for the CWB-RS to constrain and minimize exploration risks and costs \citep{steccanella2020hierarchical, Nachum2018EfficientHRL} while providing direct control and interpretability to the educators  \citep{shu2017hierarchical, lyu2019sdrl}. Ultimately, under the direction of  learning objectives and strategies, the set of problems that the CWB-RS will have to solve would be limited to balancing reranking requests from different active strategies and prioritizing one objective over the few others defined in the current high-level learning strategy.

\subsubsection{Noisy Rewards and Action Results}
An additional constraint comes from the difficulty of characterizing the toxicity of the social media content (See \ref{sec:ThreatDetection}) on which the RS must act. This results both in erroneous recommendations (e.g. content that was mistakenly supposed to be toxic undergoes reduced propagation speed) and  stochastic rewards (toxic content is evaluated by error as healthy and a positive reward is provided to the CWB-RS from the CE and CS estimation). While the RL method accounts for noisy actions' results, they still affect the performance of the system, both in terms of execution and learning time. Regarding noisy rewards,  literature has only recently started to provide solutions \citep{wang2020reinforcement,huang2019deceptive}. Still, it must be noted that in our setting, getting a positive reward for something that was considered positive should not crucially impair the acquired RS policy as the system allowed the propagation of something that it evaluated healthy (or toxic) and accordingly  evaluated its reception by other SM users. Thus, in this case, the two errors may cancel each other out and take advantage of improvements in the detectors. 
Moreover, when applying RL for ITS, an additional strategy that can be leveraged  to counter these issues is to use more reliable tests that would allow for  evaluating  the state of the users and provide more reliable rewards. 
Due to social media complexities, the  effects of detectors' failures on the performance of CWB-RS can be heavy, with backfiring as the worse-case scenario.  Extensive tests would be necessary both in simulation (e.g. \citep{https://doi.org/10.1111/bjso.12286} and real-life as well as comparisons with classical recommender systems for social media that are not sensitive to content toxicity.

\subsection{Threat Detectors and Content Analyzers} \label{sec:ThreatDetection}
Social media threat detectors and Content Analyzers have multiple roles in the platform already described in Section \ref{sec:cwb-rs}.
Given the importance of social media threats, as described in Section \ref{sec:sm_threads}, researchers have been studying how to automatically identify them (some examples can be seen in Table \ref{tab:detectors}). 
Several shared tasks have been proposed and each year they become more challenging. Moreover, new evaluation criteria, such as multilingual detection at Task 5 in Semeval 2019 \citep{basile2019semeval}, different domains at HaSpeeDe in Evalita 2020 \citep{sanguinetti2020haspeede,hoffmann2020ur}, detections at the spam level at Task 5 in Semeval-2021 \citep{pavlopoulos2021semeval}, and generalisation to social media platforms other than those used in training at EXIST in IberLEF 2021\citep{exist2021}, have been included in the datasets.

Those detectors are usually defined as a classification task commonly solved using deep learning.  
Different features are used as parameters for the models. For example, in fake news identification, \cite{hessel2019something} explored the combination of different models and features, including hand-designed features, word embeddings, ratings, number of comments and structural aspects of discussion trees. 
In addition, 
another key element of the detectors is the datasets. For some threats (e.g. hate speech and fake news),  few standard datasets target social media, but that is not the case for all  threats. For violent content detection, for example, there is not a standard dataset focused on SM to the best of our knowledge. In order to overcome these limitations, works such as \citet{bilinski2016human, zhou2018violence} use a proxy dataset, such as Hockey Violence Dataset \citep{nievas2011violence}. 

Regarding the content analysis to extract users' affective state, beliefs and opinions, similar approaches are viable. 
Affective Computing aims to recognize, infer and interpret human emotions \citep{poria2017review}, distinguishing between sentiment analysis,  polarity of content (e.g. \cite{gupta2018attention, liu2017sphereface, guo2018group}), and  recognition of the emotions present in a piece of information (e.g. \cite{baziotis2018ntua, ahmad2020borrow}). In comparison, Opinion Extraction aims at discovering users' interests and their corresponding opinions \citep{wang2019survey}.
In general, the systems extract the entity or the target, the aspect of the entity,  the opinion holder,   the time when the opinion was expressed, and the opinion \citep{liu2012sentiment}.
Similarly, the  positive aspects of social media interaction, crucial for estimating the CWB, could be extracted. Still, they  have attracted less attention, but see \citep{wang2014effects, chen2017building}.

Despite the success achieved by these efforts, the robustness of these systems  is still limited.  For instance, seldom they can  generalize to new datasets and  resist  attacks (for example, word injection) \citep{grondahl2018all,hosseini2017deceiving}. 
An example of that is the case that occurred in the OffensEval shared task \citep{zampieri2019semeval}, where different hate speech classification models were compared in different subtasks. 
The best system in Subtask B (i.e. \cite{han2019jhan014}) ranked the 76th position in Subtask A that is a general and simple case of Subtask B.\footnote{We highlight that, despite this extreme case, systems tended to maintain a similar performance across the different subtasks.} This example stresses how small changes in these tasks may drastically  impact system performance informing on the challenge of applying these approaches in the dynamic contexts of  social media. Some recent models can generalise the task while maintaining similar results in different platforms and languages under certain conditions \citep{wilkens2021mb}. 


\section{Use Case} \label{sec:use_case}
The following scenario is an example of how the Companion enables the personalization of educational interventions  to help develop users' resilience against social media threats. The focus of this use case scenario is on the algorithmic threat of filter bubbles and how it can affect the users' perspective of healthiness. The content threat is associated with body image concerns \citep{marengo2018highly}. 

Alex is a 15-year-old high school student who spends a fair amount of his free time on his phone on a daily basis.  

\emph{Without the Companion:} Alex scrolls through his social network newsfeed and encounters a photo of an influencer that promotes masculinity. As summer is approaching, he decides to check the influencer's profile for possible tips to help him tone his body. Alex spends the next hour watching videos in the influencer's profile and starts following similar profiles. The social media platform algorithms learn that Alex is interested in posts related to masculinity, and he can spend hours interacting with this type of content. Thus, to maximize  engagement, the platform starts displaying more content related to masculinity. Occasionally, the platform presents an advertisement in the form of a post to indulge Alex to buy a related product. Alex now finds his newsfeed to be filled up with fitness influencers and fitness products. Day by day, he likes and follows more fitness influencers, slowly leading  his newsfeed to be full of fitness influencers that promote a specific body type. Through time, Alex's opinion regarding beauty standards starts to shift. He starts to believe that the male body needs to be muscular to be considered attractive and healthy. When looking in the mirror, he now feels that his body is far away from being considered attractive, and he will never be able to reach the beauty standards that have been set. He starts feeling unhappy with his body and seeks comfort through his social media platform. He comes across an influencer that promotes a product for rapid muscle growth and decides to look further into his profile. There he encounters photos that show a drastic change in the influencer's physical appearance  claimed to be the result of the product. Alex decides that this product is the solution to his problem and buys it. 

\begin{figure}[h] 
\caption{\emph{Educational strategy example.} A visual example of how the policy to improve body shape-related behaviour is accomplished within the platform. An initial questionnaire is completed by the user to determine if their behaviour is classified as healthy or toxic. In the scenario that the questionnaire results come back as healthy, the user is placed into a free social media navigation state. This state will be terminated when the system detects that the user’s behaviour is no longer classified as healthy. This classification is done by analysing the profiles the user has been following based on their category and further analysing them with image classifiers. In this case, the system detects that the user's behaviour has shifted from healthy to toxic a learning activity is initiated. The user is then placed into a state where the system alters the content they receive in their newsfeed.}
\centering
\includegraphics[width=1.0\textwidth]{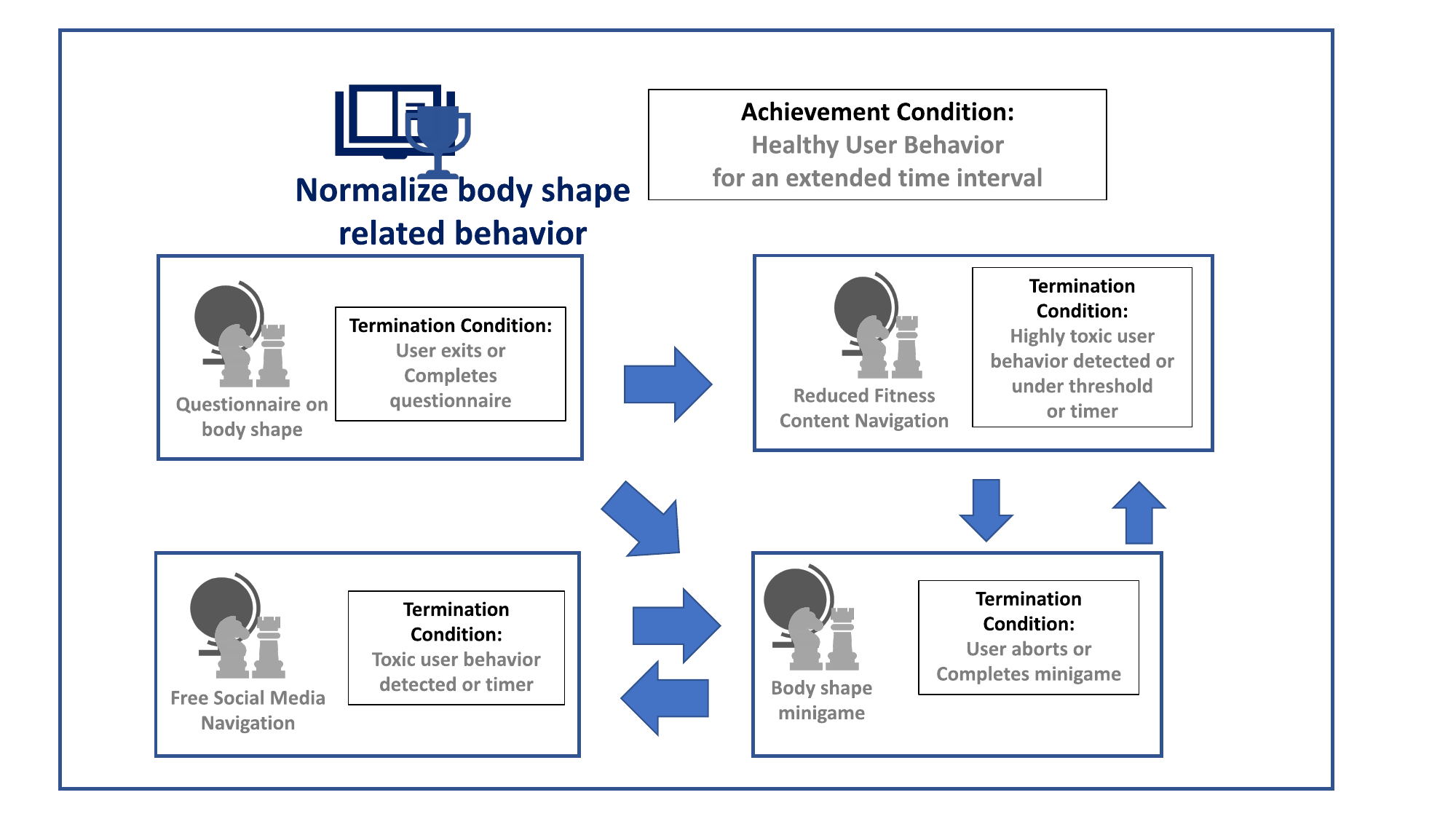}
\label{img:objective_policy}
\end{figure}

\begin{figure}[h]
\caption{\emph{A visualization of the hierarchical structure of the educational strategy}. Each educational strategy (narrative script) has a set of educational objectives that can be reached by a sequence of adaptive learning activities. The learning activities can be in the form of free-roaming, guided roaming, quizzes, minigames, or participating in group tasks. They are triggered based on the user's behaviour within the platform.}
\centering
\includegraphics[width=1.0\textwidth]{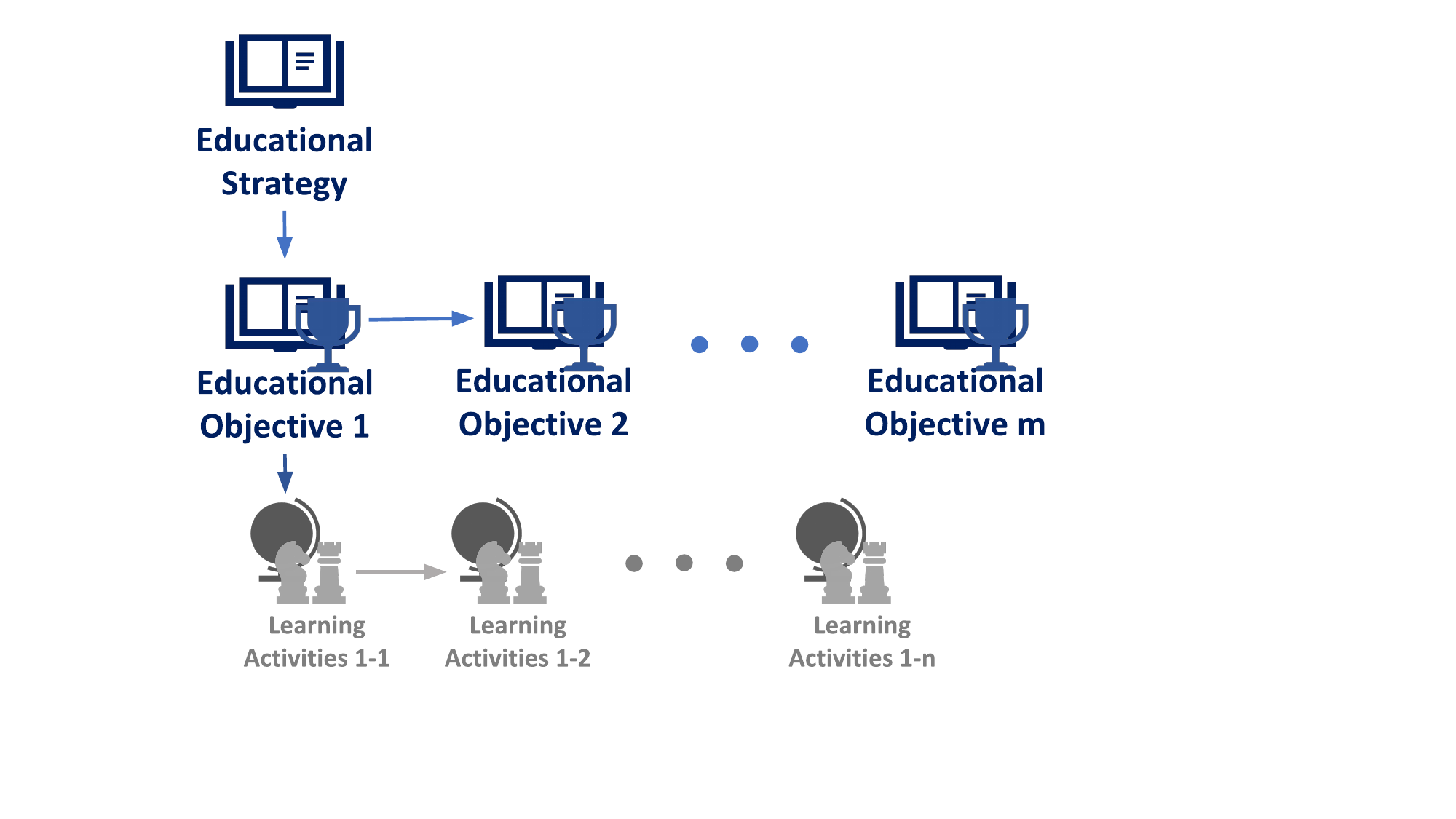}
 \label{img:edu_strategy}
\end{figure}

\emph{With the  Companion:} Alex scrolls through his social network newsfeed and encounters a photo of an influencer that promotes masculinity. As summer is approaching, he decides to check the influencer's profile for possible tips to help him tone his body. Alex spends the next hour watching videos in the influencer's profile and starts following similar profiles. The  Companion runs in the background and detects that the majority of profiles Alex has started to follow fall under the category of fitness. Image classifiers further identify that those profiles promote a specific body type. Then, the Companion triggers a narrative script and notifies Alex that a new game (the script) is available  (Figure \ref{img:objective_policy}). Alex accesses the game and initiates the narrative script. The narrative script mechanisms assign him to an influencer that supports the opposite perspective (counter-narrative) than the one he triggered. He is instructed to navigate through the profile and self-reflect on how this profile makes him feel. Alex is asked to participate in an online collaborative game showing the impact of social media influence and filter bubbles on our decision capabilities (e.g. see \citep{LomonacoOgnibeneTrianniTaibiHelmeto2022} and section \ref{studiesresults}). He is then shown a brief video of how SM algorithms work and how they can place a user into filter bubbles. In the next screen, Alex enters a mini-game where he is instructed to manipulate a filter bubble by following and unfollowing profiles and by liking and unliking posts. During the game, Alex can see how the newsfeed of the user changes according to his behaviour. Alex starts to understand how social media works and how algorithms can learn from our behaviour. Once the game is over, the narrative script ends, and Alex receives a badge for completing it. The educational component registers Alex's signs of progress and  marks the learning objective of filter bubbles as complete (Figure \ref{img:edu_strategy}). Alex returns to his social media profile and receives a notification from the Companion that the content of his newsfeed has been altered by the CWB-RS component to reduce the harmful content that he has been receiving. He has the option to revert this setting, but he decides to continue with it. The CWB-RS component filters Alex's news feed with images unrelated to muscular fitness. Eventually, this alters Alex’s content needs and influences him to start following profiles that are not solely related to muscular fitness, which leads to minimising his exposure to influencers promoting a perfect body. To confirm that Alex is staying on the right track, a few days later, the Companion operates a further inspection to analyse the content being followed. The Companion verifies that  Alex’s online behaviour  has improved after the completion of the mini-game and it does not trigger any further mini-games for him. Alex receives a notification informing him that the CWB-RS component has stopped altering his newsfeed. His newsfeed content has now become more balanced. Alex has become less obsessed with the idea of having a muscular body.

\section{Preliminary Experimental Results}
The realisation of the COURAGE  companion is progressing through the study of different educational strategies and the development and testing of educational tools and computational components. 
\subsection{Educational and Psychological Studies}
\label{studiesresults}

Educational and psychological studies are the starting point to define objectives, methodologies, and tools that will be integrated into the Companion and guide the participatory design  of the \textit{educationally managed social media community.}

Data was collected through online studies to calculate correlations between toxic content  tagging (e.g., disagreement measure) and personality traits (e.g., cognitive empathy or authoritarianism).   A link between learners' judgments and personality traits could only be weakly found for authoritarianism \citep{aprin22}.
We also studied users' intentions to share emotional images. The study was conducted in Italy and Germany, with university students surveyed online in Germany. The evaluation of nearly 200 students is not yet completed. It is expected that results will provide insights into the relationships between socio-emotional competencies, moral values, and the willingness to share images with diverse social groups. 

Similarly, a pilot study aimed to investigate the relationships between emotional intelligence and social media threats was conducted involving 110 adolescents of a secondary school in Italy during an extracurricular school activity \citep{Scifo2022Adolescents}. In particular, two research studies have been conducted within this pilot. The first study had the purpose of investigating the relationships between emotional intelligence and adolescents’ ability to detect fake news on social media. The second study included a training path aimed at stimulating emotional intelligence and promoting a conscious use of social media. Moreover, the training path has also contributed to raising adolescents’ awareness of bullying and cyberbullying. The analysis of the results is ongoing. These studies will drive the development of new educational components for the companion and help to define the companion's personalized educational strategies.

We tested a game-based educational experience \citep{de2014serious} to increase students' awareness of social media algorithmic threats, focusing on filter bubbles and echo chambers inspired by the ``wisdom of crowds'' \citep{lorenz2011social,becker2017network}. It was tested with both University and High school students providing encouraging results \citep{LomonacoOgnibeneTrianniTaibiHelmeto2022}. While more data is being collected a specific component is being designed to reproduce the experience inside the companion.

Furthermore, we developed a scenario to inform students about  racist content on social media. Here, users are informed about the background of racism by the virtual companion in a closed social media environment. By means of the results of an experimental study in which the virtual learning companion either transmits information on racism (experimental group) or not (control group), we will analyse the effects on users’ knowledge and awareness regarding racism. 

Also, we are constantly working on a scenario for  empathy training which shall sensitize young users regarding the negative effects of cyberbullying. For this training, for instance, a video showing an example and providing a definition of empathy will be shown to students in an experimental study in Germany and Spain. We hypothesize that students who completed the empathy training will be more sensitive to cyberbullying and are less likely to intend to bully in the future.

Finally, in \citep{taibi2021innovative} we present a platform specifically designed to support the development of competences related to Information and Data Literacy. This platform extends the open-source alternative to Instagram called Pixelfed, with functionalities designed to support students in increasing their awareness of social media mechanisms based upon  artificial intelligence algorithms. A pilot with secondary school students has been conducted to experiment educational activities based on the proposed platform.

\subsection{Educational Components}
\label{educational_comp_tests}

Several components and applications, that will later be integrated in the Companion, are being developed and tested.

A number of mini-games to increase social media awareness, covering topics such as the digital footprint, social media addiction, misinformation and body image dissatisfaction have been designed and tested. For instance, the serious mini-game ``SwipeIt" for sensitizing students to toxic content (e.g., cyberbullying), was  endowed with additional features like a multi-language interface. 




One of the scenarios using the  Companion  aims at raising learners' awareness of fake content in an Instagram-like  social media environment \citep{aprin22}. The VLC guides the learners through various examples in a chatbot-like dialogue. Additionally, learners are provided with access to other instances of the embedded images that are found through Google Reverse Image Search. The idea is that seeing the image in other contexts provides clues for judging the credibility of the presented content. The Companion in this scenario has been implemented as a Chrome browser plugin, which allows for running the scenario in a familiar web environment.  Initial tests with a heterogeneous group of users indicated that the environment is perceived as supportive and usable for the classification task. Subsequently,  the scenario has been tested in a secondary school classroom setting with 30 students. Preliminary findings suggest that the Companion was effective in supporting the decision about the veracity of the images shown.


The Narrative Scripts for empowering digital and self-protection skills of users through the use of computer-supported collaborative learning activities and the help of a virtual companion were presented at the Sixteenth European Conference on Technology Enhanced Learning in Italy \citep{hernandez2021narrative}. 
Over 200 school workshops were conducted involving over 1.000 adolescents in private and public schools in Barcelona. Simultaneously, a light version of school workshops and the study was replicated at the University of Campo Grande (Brazil). These workshops contributed to the testing and the fine-tuning of the educational tools developed by UPF. Data collection included information derived from the implementation of Narrative Scripts, PyramidApp and EthicsApp, based on the studies Collaborative Learning for digital Environment (CSCL), Sequencing in Learning, and the evaluation of Narrative Scripts to raise teens’ social media awareness.

Most workshops have been media education interventions with Narrative Scripts. The final result consisted of a social media simulated environment supported in Pixelfed. The pilots consisted of a six-module intervention with teaching and learning activities supported by Narrative Scripts and other gamifying elements. The interventions were diversified to integrate interactive features supported by AI elements, image decorations and “smart narratives” (allocation of roles/counternarratives; decorations in shared content). The first results of the data collected evaluate how the adaptive educational intervention embedded in the Narrative Scripts facilitates a suitable approach to educating adolescents about body image and stereotyping in social media. In particular, the analysis examines and compares approaches to identify the dominant body image stereotype in students’ social media. Results showed that the use of xAPI (tracking user behaviour in Pixelfed) combined with self-reported answers can provide a satisfactory detection of adolescents' educational needs, so as to enable automatic distribution of suitable counter-narratives (out of a collection) to students in the scripts \citep{hernandez2022narrative}

We also developed a visual interface that augments tweets with machine learning-based detectors of different forms of toxic content. To help the interpretation of this information created by state-of-the-art components, a web page was built showing correct and erroneous results produced by the detectors on different types of content, that will soon be tested in educational activities in high schools.

\subsection{Computational Components}
\label{detect_results}
The computational backbone of the Companion, which comprises diverse components such as content popularity predictors, user models and recommenders, is being developed, tested and outlined in \citep{helmeto_2022_technical_abstract}. Particular effort has been devoted to the development of content-based threat detectors because of their multiple roles: a) triggering specific educational activity, b) evaluating community well-being, and c) supporting recommendation and re-ranking of content.

Models to detect fake news and irony were presented at LREC 2022 \citep{hartl_kruschwitz_2022,turban_kruschwitz_2022}. The fake news detection system has established a new-state-of-the-art benchmark performance on the commonly used FakeNewsNet dataset. To improve performance and find the best trade-off with computational cost,  the detectors were continuously updated and  different architectural patterns (e.g. graph neural networks) were explored. The results were presented at several competitions about fake news detection as well as topics around hate speech \citep{wilkens2021mb,wilkens2021bicourage,lomonaco2022courage}, organized within the scope of well-established annual events such as CLEF 2021, CLEF 2022 and GermEval 2021. Although submissions were very competitive, the contributions by UR resulted in winning the German cross-lingual fake news detection challenge at CLEF 2022 “CheckThat!” \citep{udo2022clef} and being runner-up in the fact-claiming comment identification at GermEval 2021 \citep{tran_kruschwitz_2021_Germeval}. 

\begin{figure}
\caption{\emph{Simulated impact of different recommender systems on users' satisfaction and content diversity exposure.} Satisfaction is assumed as a proxy for the sustainability of the social media platform. Content diversity exposure could play an important role in countering the effects of filter bubbles \citep{nikolov2015measuring, bozdag2015breaking}, echo chambers \citep{wolfowicz2015social,bessi2016personality,gillani2018me}, and ultimately society polarization. Mean and standard deviation over 10 runs  with three different new connections \textit{Recommenders}: maximize opinion diversity, random, overlapping neighbourhood. For each strategy (colors) Satisfaction (‘o’) and diversity (‘x’) are pictured. \textit{Overlapping}: recommend users with the highest  number of common friends. \textit{Diversified}: recommend users with the highest opinion difference. \textit{Random}: baseline, recommend random users. 
\textit{Satisfaction}: the mean distance for each user between his opinion and the ones in his feed. \textit{Diversity}: entropy of binned opinions that populate users’ feed in each time step.  
Highlighted areas represent standard deviation across different runs. 
Each social network is initialized with 100 users (nodes) and connections (edges) are created with an adaption of preferential attachments \citep{albert2002statistical}. Differently from \citep{albert2002statistical} the nodes' probability of  being connected with an incoming node is not proportionally related to nodes'  degree but is related with their opinion distance.
Users were modelled by extending the model proposed in \citep{https://doi.org/10.1111/bjso.12286} with a backfiring component \citep{backfire1}, i.e. users exposed to opinions that were distant from theirs moved in the opposite direction. The recommender that maximizes diversity between the pairs of users to connect showed a slower start but achieved higher exposition to more diverse content and a similar level of satisfaction to the other two RSs. }
\label{fig:RSs_diversity_satisfaction}
    \centering
    \includegraphics[width=1.0\textwidth]{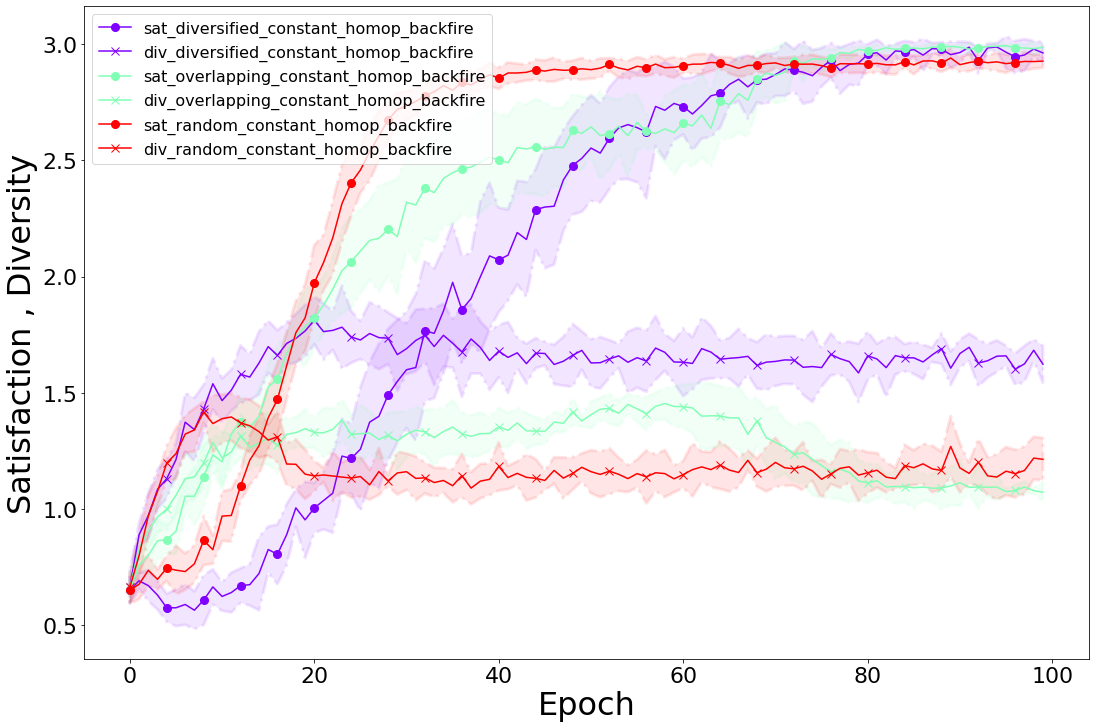}
\end{figure}

Finally, we experimented with different models of social network connectivity and user behaviour. Several computational experiments showed that recommender systems have a substantial impact on the user experience on social media. For example, we simulated the impact of different recommender systems on combinations of users' satisfaction and content diversity exposure as proxies of potential components of the CWB metrics. Satisfaction is assumed as a proxy for the sustainability of the social media platform. Content diversity exposure could play an important role in countering the effects of filter bubbles \citep{nikolov2015measuring, bozdag2015breaking}, echo chambers \citep{wolfowicz2015social,bessi2016personality,gillani2018me}, and ultimately society polarization.  In the results shown in  Fig. \ref{fig:RSs_diversity_satisfaction}. We compare three different  new connection recommenders: maximize opinion diversity, random, overlapping third order neighbourhood. Users were modelled by extending the model proposed in \citep{https://doi.org/10.1111/bjso.12286} with a backfiring component \citep{backfire1}, i.e. users exposed to content presenting opinions distant from theirs changed their minds in the opposite direction. 
The recommender that maximizes the diversity of opinion between the pairs of users to connect showed a slower start but achieved higher exposition to more diverse content and a similar level of satisfaction to the other two RSs. In the near future, we aim at integrating a full CWB-RS with educational objectives in the simulation.

\section{Discussion and Conclusion}
      This contribution is motivated by the desire to improve the  impact of social media on our society. 
      They have indeed several positive effects \citep{wang2014effects,chen2017building}:
      they extend our capacity to be  connected with our contacts, create new useful social connections, and scale up and accelerate social interactions. 
      Moreover, they   supported various forms of activism \citep{gretzel2017social,murphy2017dawning} and even enabled   whistle-blowing in oppressive regimes  \citep{joseph2012social} as well as  protests organization \citep{gladwell2011innovation,shirky2011political}.
      However, what can be defined as an explosion of SM  has also brought several new negative social phenomena, such as digital addiction \citep{YOUNG2017229,kuss2011online} and exacerbated existing ones, e.g. misinformation (wildfires) \citep{Webb2016},  which existed only on a limited scale and slow pace before. 

Teenagers are a group that is  particularly affected by numerous social media threats \citep{clarke2009early,ozimek2017materialists}. 
 We propose an educational and support platform, a Companion, focused on rising teenagers'  `new media literacy' \citep{scolari2018transmedia},  `digital citizenship' \citep{jones2016defining,xu2019social},   and awareness of social media threats. 
 The Companion will allow the smooth passage from everyday life use of social media to an educational experience by interfacing with the students to support and guide their interaction with the social media environment both inside and outside the classroom. Several  components of the Companion have been developed and successfully tested, as briefly described in \ref{educational_comp_tests}.

In social media communities, as in any society,   the safety and well-being of its members  are determined by their own mutual interactions \citep{jones2016defining}. 
%
Therefore, an important endeavour is to increase users' awareness of  the consequences of their actions and acceptance of necessary boundaries, especially in such deindividuating  environments \citep{lowry2016adults}.
 The presence of a trade-off between users' rights and duties or freedom VS safety introduces ethical issues \citep{UEAIETHICS,ienca20} (e.g. defining what is considered hate speech) that require the formulation of a comprehensive and shared view of the values of the social media community. 
 This led to the introduction of  the concept of Collective Well-Being (CWB) for Social Media communities, the shared view of the desirability of the conditions of the specific community, which would drive the definition of the educational objectives and the desired behaviours of the  community members.
 %
 To define the desired social media community as well as  the corresponding CWB objectives,  the explicit community regulations, and  the educational objectives necessary to support them, we argued for a collaborative participatory design approach  involving experts, educators, and community members, i.e. parents and teenagers \citep{Sanchez-Reina2022}. 

    In sec. \ref{sec:cwb_cc_cs_ce} a methodology is proposed to measure from online behaviour the CWB of social media communities.
Defining an operational measure of CWB could help deal with the cognitive and algorithmic threats that characterize social media and may hinder the effectiveness of purely educational efforts.
 A CWB measure could help transfer the community interests and values, as well as the educational objectives,  to the recommendation algorithms that drive the users' experience by selecting and ordering feeds and connections. 
       %
In the Companion this will be realized by the Collective Well-Being Recommender System (CWB-RS), which sequences educational activities and balances the content presented to the students in order to maximize the CWB (see sec. \ref{sec:cwb-rs}).


From a technical point of view, the problems are multiple. 
Starting from the formulation of the CWB measure, the number of aspects to balance and the likely non-linear interactions between the single and the community sub-groups will require  an iterative design approach. Moreover, while the state of the art for  the components that detect the relevant quantities is constantly improving (e.g. sharing of hate speech in the community, see sections \ref{detect_results}, \ref{tab:detectors}, and \ref{sec:ThreatDetection}), the process is still noisy. The development of active evaluation methodologies, possibly involving  educators as humans-in-the-loop, is a possible way forward.
%
The CWB-RS must face  additional complexities to evaluate the longer-term impact of its recommendations for the achievement of educational objectives and the future CWB of the community as well as balancing the level of engagement necessary for the educational and  social functions (e.g. finding out that a friend  needs online support) while avoiding digital addiction. We discussed in section \ref{sec:cwb-rs} that these issues may require combining an intelligent tutor system  with  recommender systems built using the  hierarchical reinforcement learning framework.
    Our contribution in this paper, in particular our experimental studies,  are specifically designed for relatively small communities that can tailor the approach to their own specific needs.      
    One may be tempted to think about scaling up the whole approach and integrating the educator in the loop, the CWB and CWBRS on the global social media platforms.
    This would prohibitively escalate the moderation costs that are already very demanding \citep{steiger2021psychological} and would  have to take also the educational aspect into account, which requires wider expertise and user-specific policies.
    %
      %
From an ethical point of view, the undertaking would be enormous. While  privacy, censorship,  freedom of speech, misinformation campaigns and hate speech are strongly involved ethical problems, they are by now very common in the discussion about social media \citep{Webb2016}, especially after Twitter permanently banned Donald Trump \citep{Courty2021}. However, the formulation of a CWB for social media  requires not only  formulating a metric that  balances many different demands but it  justifying the worldwide and cross-cultural adoption of a value set that supports  such a metric applied to the social and dynamic version of the \textit{WWW}. 
          
Currently, the international community is undertaking a substantial effort in  understanding and regulate the ethical implication of AI systems \citep{UNESCOAIETHICS,OECDAIETHICS,UEAIETHICS,IEEEAIETHICS}. Unmistakably there are  mixed ethical and technical issues that go beyond those currently faced  \citep{UEAIETHICS,IEEEAIETHICS,OECDAIETHICS,UNESCOAIETHICS}\footnote{see \href{https://standards.ieee.org/standard/7010-2020.html}{IEEE 7010-2020 - IEEE Recommended Practice for Assessing the Impact of Autonomous and Intelligent Systems on Human Well-Being}}. For example, trying to optimize the CWB may induce a further increase in  social media complexity. This may  reduce even more our control over social dynamics  \citep{Floridi2014-FLOTFR-3} and backfire with even more threatening, addictive, and unhealthy dystopian situations. 

     While it is crucial that the international community continues its effort and targets social media \citep{gorwa2019platform},  
      we highlighted that aiming to improve the CWB of SM local communities implies first and foremost aiming to educate local  communities themselves, as the CWB depends on users' attitude, interactions and relationships  \citep{jones2016defining}. 
      Education is the best way we know to improve human behaviour. 
    Indeed, if the methodology is successfully applied on a sufficient scale improving the members' new media literacy and  digital citizenship, it may improve the general impact of social media on our society.
        Focusing on more controlled communities, e.g. schools, with  a very  limited scale for the social media domain reduces the ethical burden on the design side as well as the technical demands for accuracy and reliability through the integration of a mediator role for educators and parents,  through a ``Human in the Loop'' paradigm. This approach also allows focusing on the critical educational aspect. The creation of educationally managed social media communities allows  supported learning experiences and a full range of new experiments \citep{Amarasinghe2021computersupportedcollaborationscripts,HernandezLeo2021NarrativeScripts, Fulantelli2021SchoolPromoteConsciousSocialMedia,
    Malzahn2021Disagreement}.  
    
    Differently from previous other interventions with a similar aim, this paradigm enabled by the educational virtual Companion for social media has indeed the potential to provide an educational experience on a scale comparable to that of the social media platforms. 
    Indeed, it will be challenging to define an educational path that covers most of the numerous points of interest in digital citizenship \citep{jones2016defining,xu2019social}.    
    Also, the integration of this educational experience in student life is challenging, especially regarding the  experience outside the classroom, where the non-educational global  platforms will compete for student time and attention.  
    Still, we believe that the combined technological and educational strategy implemented by the  Companion has a good chances to be effective in containing many of the current social media threats.

        Finally, this approach is the perfect means to bootstrap and test the concept of CWB-RS systems, verify their feasibility, stability and robustness, and create suitable datasets. The data collected from this initiative may not only be useful for replicating and extending this type of educational approach, but it could also be a first step to provide evidence that social media's impact on society can be improved by taking the community needs more into account in their design. A characteristic feature of social media is that they are crucially the result of a community activity, which both consumes and produces their content. It is daunting that platforms' objectives are so detached from those of the community.
        Therefore, we hope that our results can  support the process of introducing new evidence-based  regulations both for the platforms and their algorithms, beginning with requesting the  platforms to release their  data for scientific research and  enable  large-scale studies, which have been curbed after the limits they recently set  following Cambridge Analytica and other scandals \citep{Hemsley2019}.

    \newpage
\begin{table}[]
\centering
{\renewcommand{\arraystretch}{1.6}
\begin{tabular}{|m{3cm}|p{9cm}|p{4cm}|}
\hline
\emph{Categories}  & \emph{Description}& \emph{Reference} \\ \hline
\emph{Vitality} &``The vitality domain includes... emotional health, with positive and negative affect, optimism and emotional intelligence.'' 
& \citep{hong2017user} \\ \hline
\emph{Opportunity} & the ``perceived opportunity to achieve life goals and socioeconomic mobility’’, ``influenced by ... access to education and training’’&  \\ \hline
\emph{Connectedness} &``The connectedness domain assesses the level of connection and support among community members... Human relationships and relatedness are fundamental for the achievement of well-being according to many foundational theories of well-being.\dots Connectedness includes dimensions of social acceptance (i.e., positive attitudes toward people) and social integration (i.e., feeling a sense of belonging to the community).’’ 
& \citep{seligman2011flourish, fredrickson2004broaden,cohen1985stress, ryff2004positive, dunn1959high, walker2011social, lopez2009oxford} 
\\
\hline
\emph{Contribution} &``The contribution domain incorporates residents' feelings of meaning and purpose attributed to community engagement and belonging (e.g. volunteering, civic engagement, or belonging to a religious or community group). Sense of purpose is a cognitive process that provides personal meaning and defines life goals.’’
& \citep{roy2018collective,forgeard2011doing, keyes2012mental} \\
\hline
\emph{Inspiration} &``The inspiration domain includes community members’ perceived access to activities that are intrinsically motivating and stimulating… [such as]  life-long learning, goal-striving, creativity, and intrinsic motivation.’’ 
& \citep{roy2018collective,meier2018positive} \\ \hline
\end{tabular}}
\caption{Categories of properties of social media communities relevant for Collective well-being and education extracted from the framework presented in  \citep{roy2018collective}}
\label{tab:cwb-categories}
\end{table}

\begin{table}[]
{\renewcommand{\arraystretch}{1.6}
\begin{tabular}{|p{8.2cm}|p{8.2cm}|}
\hline
\multicolumn{1}{|c|}{\textit{Content Based Social Media Threats}}  & \multicolumn{1}{c|}{\textit{Social Media Cognitive and Socioemotional  Threats}}  \\ 
\hline
    \begin{tabular}[t]{@{}l@{}}Toxic Content \citep{Kozyreva2020Citizens}          \\ Fake news/disinformation \\ \citep{de2018multi}        \\ Bullying \citep{grigg2010cyber,Mladenovic2021}                     \\ Hate speech \citep{zimmerman2018improving}                             \\ Stalking \citep{tartari2015benefits}                \\ Discrimination \citep{10.1145/3178876.3186140}      \\ Radicalization \citep{johnson2016new}          \\ Smoke \citep{christakis2008collective}              \\ Sexism/sexual harassment \citep{barak2005sexual}    \\ Objectification \citep{ozimek2017materialists}      \\ Beauty stereotypes \citep{verrastro2020fear}\end{tabular} 
& 
    \begin{tabular}[t]{@{}l@{}}Impulsivity \citep{Lee2019a} \\  Fear of Missing Out 
    \citep{Alutaybi2019} \\  Confirmation bias \citep{del2017modeling}\\\citep{knobloch2012preelection} \\  Social reinforcement \citep{liu2018deep} \\  Backfire effect \citep{backfire1} \\  Attention limit \citep{weng2012competition} \\  Emotional load \citep{kramer2014experimental}\\\citep{brady2017emotion} \\  Anonymity \citep{urena2019review} \\  Depersonalisation \citep{diener1980deindividuation}\\\citep{postmes1998deindividuation} \\  Digital addiction \citep{pmid25426088} \\ 
    \citep{kuss2011online,almourad2020defining} \\   Lack of digital literacy \citep{xu2019social}\\ \citep{whittaker2015cyberbullying}\end{tabular} \\
\hline \hline 
\multicolumn{1}{|c|}{\textit{Social Media Dynamics induced Threats}} & \multicolumn{1}{c|}{\textit{Algorithmic Social Media Threats}} \\ \hline
    \begin{tabular}[t]{@{}l@{}}Filter bubbles \citep{bozdag2015breaking}\\ \citep{nikolov2015measuring,https://doi.org/10.1111/bjso.12286} \\ Echo chambers \citep{gillani2018me} \\ Digital wildfire \cite{Webb2016}\end{tabular}
& 
    \begin{tabular}[t]{@{}l@{}}Content diversity \citep{doi:10.1287/isre.2013.0497} \\  Misclassification \citep{stocker2020riding} \\  Algorithmic bias \citep{chen2020bias} \\  Malicious users \citep{zhou2017proguard} \\  Gerrymandering \citep{stewart2019information}\end{tabular}  \\
\hline
\end{tabular}}
\caption{Examples of social media threats distinguished into three categories (content; algorithmic; network, and attacks; and dynamics) and examples of cognitive phenomena that may exasperate them.}
\label{tab:sm_threats}
\end{table}

\begin{table}[]
\centering
{\renewcommand{\arraystretch}{1.6}
\begin{tabular}{|p{5.0cm}|p{11.4cm}|}
\hline
\emph{Type of detector}  & \emph{Reference} \\ \hline
Stance detection & \citep{zarrella2016mitre, augenstein2016stance} \\
Controversy identification  & \citep{hessel2019something, zhong2020integrating} \\ 
Fact-checking & \citep{dale2017nlp, wang2017liar, long2017fake, atanasova-etal-2020-generating, liu2019text, nie2019combining, jobanputra2019unsupervised}\\ Hate speech & \citep{indurthi2019fermi, cer2018universal, basile2019semeval, nikolov2019nikolov}\\
Violence recognition & \citep{bilinski2016human, perronnin2010improving, zhou2018violence, nievas2011violence}\\
Gender bias & \citep{Prost2019}\\
Offensive content & \citep{zampieri2019semeval,hosseini2017deceiving}\\

\hline
\end{tabular}}
\caption{Short list of works on social media threat detection and content analysis exemplifying the variety of approaches and works.}
\label{tab:detectors}
\end{table}

\section*{Conflict of Interest Statement}

The authors declare that the research was conducted in the absence of any commercial or financial relationships that could be construed as a potential conflict of interest.

\section*{Author Contributions}

Dimitri Ognibene conceived of the presented ideas and concepts. He contributed to the design of the educational platform architecture. 

Rodrigo Souza Wilkens supported and revised the design and developed the content machine learning aspects.

Udo Kruschwitz proposed and described the integration of behavioural economics methodologies for education. He supervised the information retrieval aspect of the contribution.

Francesco Lomonaco helped with the aspects of network dynamics and dataset collection.

Davinia Hernández-Leo, J. Roberto Sánchez-Reina, Emily Theophilou, and Rene Alejandro Lobo contributed with the proposal of playful educational methodology, the cooperative social media design and narrative scripts.

Ulrich Hoppe, Nils Malzahn, Sabrina Eimler overviewed the conceptual development.

Davide Taibi and Lidia Scifo contributed with expertise on digital addiction and digital literacy intervention design.

\section*{Funding}
This work has been developed in the framework of the
project COURAGE - A social media companion safeguarding
and educating students (no. 95567), funded
by the Volkswagen Foundation in the topic Artificial
Intelligence and the Society of the Future.




\bibliographystyle{frontiersinSCNS_ENG_HUMS} 
\bibliography{allreferences}

\end{document}